\documentclass[12pt]{article}
\pdfoutput=1
\usepackage{jheppub_X}

\usepackage[utf8]{inputenc}

\usepackage{cleveref}
\crefname{table}{Table}{Tables}
\crefname{equation}{Eq.}{Eqs.}
\crefname{appendix}{App.}{Apps.}
\crefname{section}{Sec.}{Secs.}
\crefname{figure}{Fig.}{Figs.}
\usepackage{amsmath}
\usepackage{setspace}
\usepackage{tikz}
\usetikzlibrary{decorations.markings}
\usepackage[many]{tcolorbox}
\usepackage{bm}
\usepackage{xfrac}
\usepackage{pbox}
\usepackage{comment}
\usepackage{graphbox}

\allowdisplaybreaks

\makeatletter
\g@addto@macro\bfseries{\boldmath}
\makeatother

\def\eg{\textit{e.g.}}
\def\ie{\textit{i.e.}}

\newcommand{\dd}{\mathrm{d}}
\newcommand{\lag}{\ensuremath{\mathcal{L}}}
\newcommand{\amp}{\ensuremath{\mathcal{A}}}
\newcommand{\canonh}{\ensuremath{\mathbf{h}}}
\newcommand{\s}{\hspace{0.8pt}}

\newcommand{\normalcoordssuffix}{\vert_{\substack{\text{normal} \\ \text{coords}}}}

\title{\Large
Unitarity Violation and the Geometry of Higgs EFTs
}

\author[a]{Timothy Cohen,}
\author[b]{Nathaniel Craig,}
\author[a]{Xiaochuan Lu,}
\author[c]{and Dave Sutherland\hspace{1pt}}

\affiliation[a]{Institute for Fundamental Science, University of Oregon, Eugene, Oregon 97403, USA}
\affiliation[b]{Department of Physics, University of California, Santa Barbara, CA 93106, USA}
\affiliation[c]{INFN Sezione di Trieste, via Bonomea 265, 34136 Trieste TS, Italy}

\emailAdd{tcohen@uoregon.edu}
\emailAdd{ncraig@ucsb.edu}
\emailAdd{xlu@uoregon.edu}
\emailAdd{dsutherl@sissa.it}

\abstract{
We derive the scale of unitarity violation from the geometry of Effective Field Theory (EFT) extensions of the Standard Model Higgs sector. The high-energy behavior of amplitudes with more than four scalar legs depends on derivatives of geometric invariants with respect to the physical Higgs field $h$, such that higher-point amplitudes begin to reconstruct the scalar manifold away from our vacuum. In theories whose low-energy limit can be described by the Higgs EFT (HEFT) but not the Standard Model EFT (SMEFT), non-analyticities in the vicinity of our vacuum limit the radius of convergence of geometric invariants, leading to unitarity violation at energies below $4\pi v$. Our results unify approaches to the HEFT/SMEFT dichotomy based on unitarity, analyticity, and geometry, and more broadly illustrate the sense in which observables probe the geometry of an EFT. Along the way, we provide novel basis-independent results for Goldstone/Higgs boson scattering amplitudes expressed in terms of geometric covariant quantities.
}

\begin{document}
\maketitle
\flushbottom
\setcounter{page}{2}
\newpage

\begin{spacing}{1.1}
\parskip=0ex

\section{Introduction}

Treating the Standard Model as an Effective Field Theory (EFT) is an invaluable way to characterize the potential impact of new physics on observables.  Generally, there are two EFT approaches to extending the Standard Model, distinguished by the linearly realized gauge symmetry: taking the gauge symmetry of the EFT to be $SU(3)_C \times SU(2)_L \times U(1)_Y$ results in the Standard Model EFT (SMEFT) \cite{Weinberg:1979sa, Buchmuller:1985jz, Leung:1984ni}, while taking only $SU(3)_C \times U(1)_{\rm em}$ yields the Higgs EFT (HEFT)\footnote{Alternately, the Higgs-Electroweak Chiral Lagrangian (EWCh$\mathcal{L}$).} \cite{Feruglio:1992wf, Bagger:1993zf, Koulovassilopoulos:1993pw}. In the former description, electroweak symmetry is then broken by the vacuum expectation value of the electroweak doublet $H$, correlating the interactions of the physical Higgs $h$ and the Goldstones $\pi_i$ that become the longitudinal modes of the $W$ and $Z$. In the latter, no relation is assumed between $h$ and the $\pi_i$, which nonlinearly realize the $SU(2)_L \times U(1)_Y$ gauge symmetry.  The parameter space of SMEFT is a subset of HEFT \cite{Alonso:2015fsp}, implying a grading of effective theories expressed in terms of HEFT: ``reducible HEFTs'' that can be equally well expressed in terms of SMEFT via field redefinitions, and ``irreducible HEFTs'' that cannot. (In what follows, to economize our language we will simply refer to ``reducible HEFT'' as SMEFT, since the two only differ by the choice of parameterization, and ``irreducible HEFT'' as HEFT.) To the extent that EFT extensions of the Standard Model now play a central role in the interpretation of experimental data, understanding the essential differences between HEFT and SMEFT -- and the physical implications of theories whose low-energy limit can only be described by the former -- is of paramount importance. Since field redefinitions can blur apparent distinctions at the level of the Lagrangian, invariant criteria are required to draw meaningful distinctions between the two EFT frameworks.

At least three distinct approaches have been taken in distinguishing HEFT from SMEFT. The first involves {\it unitarity}: as the full electroweak gauge symmetry $SU(2)_L \times U(1)_Y$ is nonlinearly realized in HEFT (with the decay constant set by the electroweak symmetry breaking scale $v$), from Naive Dimensional Analysis \cite{Weinberg:1978kz, Manohar:1983md, Georgi:1992dw} one expects these EFTs to violate unitarity at the scale $\sim 4 \pi v$. In SMEFT, on the other hand, $SU(2)_L \times U(1)_Y$ is linearly realized, and the scale of unitarity violation can be parametrically separated from $v$. The second approach involves {\it analyticity} of the Lagrangian expressed in terms of the electroweak doublet $H$: in \cite{Falkowski:2019tft} (see also~\cite{Finn:2019aip}) it was argued that HEFT arises whenever the scalar potential for $H$ is non-analytic at $H = 0$, provided such non-analyticities cannot be removed by a field redefinition. The third approach, introduced in \cite{Alonso:2015fsp, Alonso:2016oah} and further developed in \cite{Cohen:2020xca}, involves {\it geometry}: by treating the Higgs and Goldstone bosons as the coordinates on a Riemannian manifold, invariant properties of the EFT geometry can be used to distinguish between HEFT and SMEFT. In particular, the scalar manifold of a SMEFT possesses a fixed point of electroweak symmetry at which a set of curvature invariants are all finite. Moreover, in order for SMEFT to be of practical use for calculating scattering amplitudes about our physical vacuum, expansions of these curvature invariants about the fixed point need to converge at our physical vacuum.

One would expect the three approaches to be related. Indeed, the relations between $analyticity \leftrightarrow unitarity$ and $analyticity \leftrightarrow geometry$ have been established. In \cite{Falkowski:2019tft} analyticity was connected to unitarity by demonstrating that non-analyticities in the Lagrangian for $H$ at the point $H = 0$ gave rise to unitarity violation in the inelastic scattering of two Goldstones into any number of Higgs bosons by the scale $4 \pi v$, with only logarithmic sensitivity to the dimensionless coefficient of the non-analytic term. A related perspective was presented in \cite{Chang:2019vez, Abu-Ajamieh:2020yqi}, which identified the $m \rightarrow n$ scattering processes with the greatest sensitivity to unitarity violation associated with HEFT. Similarly, analyticity was connected to geometry in \cite{Cohen:2020xca}, where we showed that the analyticity criteria for SMEFT could be rephrased in a field redefinition-invariant manner in terms of curvature criteria on the EFT manifold.

However, the connection between unitarity and geometry has remained obscure. It has long been known that scattering amplitudes probe the local geometry of the scalar manifold \cite{Alvarez-Gaume:1981exa, Dixon:1989fj, Alonso:2015fsp, Nagai:2019tgi}; for example, the leading-in-energy terms in 2-to-2 scattering amplitudes involving Higgses $h$ and Goldstone bosons $\pi_i$ are proportional to sectional curvatures evaluated at our vacuum \cite{Alonso:2015fsp}. This suggests a clear connection between unitarity and geometry, but presents something of a puzzle: if scattering amplitudes probe the geometry at our vacuum, how can they possibly tell us about the properties of a far-away fixed point on the scalar manifold, which the geometric approach tells us is key to distinguishing between HEFT and SMEFT?

In this paper, we connect the geometry of the scalar manifold to the scales of unitarity violation in HEFT and SMEFT, and more broadly explore the connection between scattering amplitudes and geometry in these theories. Intuitively, although the leading-in-energy terms in 2-to-2 scattering amplitudes depend on sectional curvatures evaluated at our vacuum, the leading-in-energy terms in higher-point amplitudes depend on derivatives of the sectional curvatures with respect to $h$, such that sufficiently high-point amplitudes begin to reconstruct the geometry away from our vacuum. This completes the web of connections between approaches to the HEFT/SMEFT dichotomy based on unitarity, analyticity, and geometry, and fully illuminates the sense in which observables probe the geometry of an EFT.

The basic idea connecting geometry and unitarity is as follows: If one complexifies the physical Higgs excitation $h$, then the existence of a non-analyticity of the effective action on the complex $h$-plane implies a finite value for the radius of convergence $v_\star$ of the Taylor expansion of any HEFT curvature invariants in $h$ about $h=0$, the physical vacuum. Amplitudes for Higgs/Goldstone scattering can be written in terms of these curvature invariants, where the leading effect of adding another Higgs to a given amplitude is to add an $h$ derivative to the curvature invariants appearing therein. Therefore, by the Cauchy-Hadamard theorem, the growth of the amplitude with the number of Higgses is determined by $v_\star$, the (smallest) radius of convergence of the involved curvature invariants. This in turn determines the scale of unitarity violation to be $\sim 4\pi v_\star$.

So far, the discussion applies to both SMEFT and HEFT. To make the distinction clearer, note that if the non-analyticities are close enough to the physical vacuum -- when $v_\star\lesssim v$ -- then a SMEFT expansion performed at the electroweak symmetric point $\langle |H|^2\rangle =0$ cannot possibly encompass the physical vacuum where $\langle |H|^2\rangle=\frac12 v^2$.  This indicates that HEFT is required as an EFT description around the physical vacuum. Then the above argument also tells us that unitarity violation happens at the TeV scale: $4\pi v_\star \lesssim 4\pi v$. This general result is verified by a series of example UV theories that we know must be matched onto HEFT.  Furthermore, by summing over different Higgs multiplicities and placing a unitarity bound on an inelastic cross section, the scale of unitarity violation is then only logarithmically sensitive to the couplings in the UV theory, as was first pointed out in~\cite{Falkowski:2019tft}.

In a bit more detail, the logical flow that leads to the main result of this paper is as follows:
\begin{enumerate}
\item EFTs, such as SMEFT and HEFT, have a limited regime of validity.  One manifestation of this fact is that there will exist amplitudes $\mathcal{A}(E)$ computed within the EFT that are unbounded functions of the energy $E$. Enforcing that the theory is unitary implies a constraint $\left|\mathcal{A}\right| \lesssim 1$, which results in a finite energy range  $E\le \Lambda$ in which the EFT is valid. This scale of unitarity violation $\Lambda$ provides an upper bound on the scale at which new physics is expected to emerge in order to maintain unitarity within a more fundamental description.
\item We are interested in finding the process(es) that yield the lowest $\Lambda$, so that we can be sure that we are not pushing the EFT into regions where it does not provide a valid description.  Within HEFT, we will identify a series of amplitudes $\mathcal{A}_n(E)$ that also grow as $n$ is increased.  The lowest value of $\Lambda$ is therefore determined by the speed with which $\mathcal{A}_n$ grows as $n$ is increased.
\item One such class of amplitudes is $\mathcal{A}_n(E) = \mathcal{A} (\pi_i \pi_j h^{n-2})$, where $\pi_i$ are the electroweak Goldstone bosons, and $h$ is the Higgs boson.  When written using a geometric language, $\mathcal{A}_n$ is determined by the $n^\text{th}$ covariant derivatives of the potential and the $(n-4)^\text{th}$ covariant derivatives of the Riemann tensor, \eg~see~\cref{eq:MasterGeometricAmp,eq:TwoGoldstoneAmplitude}. These can be further written into successive partial derivatives of curvature invariants, \eg~see~\cref{eq:NablaRiemannIsPartialSectional,eq:NablaPotentialIsPartialLaplacian,eq:TwoGoldstoneAmplitudeExplicit}.
\item This allows us to apply the Cauchy-Hadamard theorem, which relates the growth of the $n^\text{th}$ derivative of a (single variable) complex function to that function's radius of convergence. This tells us that the growth of $\mathcal{A}_n$ can be determined by finding the convergence radius of the geometric quantities that define HEFT when they are viewed as functions on the complex-$h$ plane.
\item This provides a precise connection between the scale of unitarity violation and the curvature criteria governing when HEFT is required.  Taking the physical vacuum as the origin, the curvature criteria of~\cite{Cohen:2020xca} tells us that HEFT has non-analyticities at or very close to the electroweak symmetric point on the manifold.  Therefore, HEFT violates unitarity at $\Lambda \sim 4\pi v$, with only a logarithmic sensitivity to the coupling constants.
\end{enumerate}

Our results are closely related to the arguments given in Refs.~\cite{Chang:2019vez,Falkowski:2019tft}, which established a connection between the Higgs trilinear correction $\delta h^3$ and the scale of unitarity violation. The perspective presented here has the advantage of being manifestly basis-independent thanks to the use of geometric invariants (and covariant derivatives thereof) in constructing the relevant scattering amplitudes. Both this work and \cite{Chang:2019vez,Falkowski:2019tft, Abu-Ajamieh:2020yqi} highlight an important feature of unitarity violation in HEFT, namely that 2-to-2 amplitudes generally do {\it not} provide the lowest bound on the scale of unitarity violation. Rather, the strongest bounds come from higher-multiplicity amplitudes. This result has a satisfying interpretation in terms of the geometry of the scalar manifold: while 2-to-2 amplitudes probe sectional curvatures at our vacuum, only higher-point amplitudes begin to reconstruct the curvature elsewhere on the manifold.

Understanding the necessary energy growth of diverse amplitudes in HEFT brings another question into sharper focus: how might we decisively exclude (or discover) HEFT? Equivalently, {\it can we experimentally prove that electroweak symmetry is linearly realized by the known fundamental particles?} Although this is frequently {\it assumed}  in the present era, it is far from proven. In fact, it is remarkably easy for an ultraviolet theory respecting $SU(2)_L \times U(1)_Y$ to give rise to an infrared EFT in which the symmetry is only nonlinearly realized. For example, there are a plethora of perturbative extensions of the Standard Model consistent with all current data that would require HEFT to describe their low-energy imprint on the Higgs sector \cite{Banta:2021dek}. As long as this is the case, HEFT clearly remains viable and relevant to the characterization of electroweak physics. While excluding all such perturbative examples would represent considerable progress towards ruling out HEFT, it would still leave the door open for exotic or strongly coupled scenarios. A more satisfying and comprehensive verdict may come from probing the high-energy behavior of scattering amplitudes in the electroweak sector. On one hand, it is increasingly clear that probing 2-to-2 amplitudes at energies of order $\sim 4 \pi v$ is insufficient to settle the question. On the other hand, it is likely (perhaps with some reasonable physical assumptions) that probing a sufficiently comprehensive set of 2-to-few amplitudes at the same energies could prove decisive. The question has the rare appeal of being interesting whether or not future measurements continue to agree with Standard Model predictions, providing a compelling target for the LHC and future colliders.

Although we do not attempt to answer this question here, the connections we build between amplitudes and the geometry of the scalar manifold may prove useful to the endeavor. More broadly, the general results we present for HEFT amplitudes expressed in geometric language may be of much wider utility. They build on previous studies applying geometric techniques to the study of non-linear sigma models \cite{Honerkamp:1971sh, Tataru:1975ys, Alvarez-Gaume:1981exa, Alvarez-Gaume:1981exv, Gaillard:1985uh} as well as HEFT \cite{Alonso:2015fsp, Nagai:2019tgi}, SMEFT \cite{Helset:2020yio}, and their extensions \cite{Alonso:2016oah}.  We anticipate that our results will be of use for general studies of the predictions of HEFT (beyond the simple question of the scale where unitarity is violated), and may serve as a bridge between Lagrangian parameterizations and purely on-shell approaches \cite{Shadmi:2018xan, Durieux:2019eor, Durieux:2020gip}.

The rest of this paper is organized as follows:  In~\cref{sec:Geometry}, we review the geometric formulation of generic scalar field theory amplitudes.  Then in~\cref{sec:GeoHEFT}, we specialize to the case of HEFT, and provide general amplitudes for Goldstone/Higgs boson scattering.  This allows us to apply these coordinate independent results in~\cref{sec:UnitarityCutoffs} to derive a general connection between the radius of convergence and unitarity violation.  We then apply this formalism to three concrete UV models in~\cref{sec:UVCompletions} to show that HEFT indeed violates unitarity at a scale $\lesssim 4\pi v$.  Finally,~\cref{Sec:Conc} contains our conclusions, while some technical details are relegated to a set of appendices.

\section{Geometrizing Amplitudes}
\label{sec:Geometry}
In this section, we review the connection between tree-level amplitudes that result from the generic scalar EFT Lagrangian
\begin{equation}
\lag = \frac12 g_{\alpha \beta}\big(\vec \phi\,\big) \partial_\mu \phi^\alpha \partial^\mu \phi^\beta - V\big(\vec \phi\,\big) + \mathcal{O}\big(\partial^4\big) \,,
\label{eq:GenLagUnexpanded}
\end{equation}
and the geometry of the scalar manifold.
Specifically, we will show that amplitudes derived from the theory defined by~\cref{eq:GenLagUnexpanded} may be written in terms of covariant derivatives of the Riemann curvature tensor and the potential $V$. When expressed in this form, the (non-derivative) field redefinition invariance of the $S$-matrix is manifest.
Our primary goal will be to show how choosing a special basis allows amplitudes to be efficiently constructed.
This makes it straightforward to see how amplitudes are related to the geometric invariants, in such a way that elucidates their kinematic properties.

\subsection{Preliminaries}

The real fields $\vec \phi$ may be viewed as coordinates on a smooth target space manifold $M$, which are indexed $\alpha,\beta, \ldots \in \{1,2,\ldots,\dim M\}$.
We will focus on coordinate redefinitions $\vec\phi \to \vec\chi$ of the form
\begin{equation}
\vec \phi = \vec\phi( \vec\chi) \,,
\label{eq:CoordRedef}
\end{equation}
where $\vec\phi$ is a smooth function of $\vec\chi$, admitting a Taylor expansion in powers of $\chi$ coordinates.\footnote{This is a subset of the full allowed field redefinition freedom; we neglect the freedom to add local functionals of the new fields that contain derivatives, \eg~$\phi^\alpha = \phi^\alpha(\vec\chi) + \xi^\alpha_{\beta\gamma}(\vec\chi) \partial \chi^\beta \partial \chi^\gamma$.  This would generate terms in the Lagrangian beyond two derivative order, which is outside the scope assumed in the theory defined by~\cref{eq:GenLagUnexpanded}.}
Under~\cref{eq:CoordRedef}, the components of the Lagrangian in~\cref{eq:GenLagUnexpanded} transform as a scalar and symmetric 2-form on the manifold respectively:
\begin{subequations}
\begin{align}
V_\text{new}(\vec\chi) &= V\big(\vec\phi(\vec\chi)\big) \,, \\[5pt]
g_{\text{new}\; \alpha\beta}(\vec\chi) &= \frac{\partial \phi^\gamma(\vec\chi)}{\partial \chi^\alpha} \frac{\partial \phi^\delta(\vec\chi)}{\partial \chi^\beta} g_{\gamma\delta}\big(\vec\phi(\vec\chi)\big) \,.
\end{align}
\end{subequations}
In a unitary field theory, $g_{\alpha\beta}$ must be positive definite, and so it can be interpreted as a Riemannian metric on the field space manifold, thereby also imbuing it with a notion of curvature.

By way of metric and curvature conventions, for a given tensor $T_{\alpha_1 \dots \alpha_m}$ we will write the components of partial derivatives as indices following \emph{commas}, and the components of covariant derivatives as indices following \emph{semi-colons}, thus
\begin{subequations}
\begin{align}
T_{\alpha_1 \dots \alpha_m , \beta_1 \dots \beta_n} &\equiv \frac{\partial}{\partial \phi^{\beta_n}} \ldots \frac{\partial}{\partial \phi^{\beta_1}} T_{\alpha_1 \dots \alpha_m} \,, \\[8pt]
T_{\alpha_1 \dots \alpha_m ; \beta_1 \dots \beta_n} &\equiv \frac{\nabla}{\nabla \phi^{\beta_n}} \ldots \frac{\nabla}{\nabla \phi^{\beta_1}} T_{\alpha_1 \dots \alpha_m} \,.
\end{align}
\end{subequations}
Covariant derivatives evaluate to
\begin{equation}
T_{\alpha_1 \dots \alpha_m ; \beta} =  T_{\alpha_1 \dots \alpha_m,\beta} - \sum_{i=1}^m T_{\alpha_1 \dots \hat \alpha_i \rho \dots \alpha_m} \Gamma^\rho_{\alpha_i \beta} \,,
\end{equation}
where the string $\alpha_1 \dots \hat \alpha_i \rho \dots \alpha_m$ denotes the string $\alpha_1 \dots \alpha_m$ with $\alpha_i$ replaced by $\rho$, and the metric connection can be computed using
\begin{equation}
\Gamma^\rho_{\alpha\beta} = g^{\rho \gamma} \Gamma_{\gamma\alpha\beta} = g^{\rho\gamma} \frac12 \left( g_{\gamma\alpha,\beta} + g_{\gamma\beta,\alpha} - g_{\alpha\beta,\gamma} \right) \,,
\end{equation}
The components of the Riemann curvature tensor are written
\begin{equation}
R_{\alpha\beta\mu\nu} = g_{\rho\nu} \left( \Gamma^\rho_{\alpha\mu,\beta} - \Gamma^\rho_{\beta\mu,\alpha} +\Gamma^\sigma_{\alpha\mu} \Gamma^\rho_{\beta\sigma} - \Gamma^\sigma_{\beta\mu} \Gamma^\rho_{\alpha\sigma} \right) \,,
\end{equation}
and the Ricci scalar curvature is
\begin{equation}
R = g^{\alpha \mu} g^{\beta\nu} R_{\alpha\beta\mu\nu} \,.
\end{equation}
We will also use round brackets to denote symmetrization of groups of indices;
\begin{equation}
T_{(\alpha_1 \dots \alpha_m)} = \frac{1}{m!} \sum_{\sigma \in S_m} T_{\sigma\{\alpha_1 \dots \alpha_m\}} \,,
\end{equation}
is a normalized sum over permutations $\sigma \{ \alpha_1 \dots \alpha_m \}$ of the indices.

Returning to the Lagrangian in~\cref{eq:GenLagUnexpanded}, without loss of generality, we take the physical vacuum to lie at the origin $\phi^\alpha=0$, so that the potential is extremized there:\footnote{We use `vacuum' and `origin' interchangeably from here forward.}
\begin{align}
\overline V_{,\beta} = 0 \,.
\end{align}
Bars denote quantities evaluated at the origin.

\subsection{Amplitudes}

Tree-level amplitudes may be constructed by Taylor expanding the metric and potential about the vacuum
\begin{equation}
\lag = \frac12 \sum_{n=0}^\infty \frac{1}{n!} \, \overline g_{\alpha\beta,\gamma_1 \dots \gamma_n} \, \big(\partial_\mu \phi^\alpha\big) \big(\partial^\mu \phi^\beta\big) \, \phi^{\gamma_1} \ldots \phi^{\gamma_n} -  \sum_{n=0}^\infty \frac{1}{n!} \, \overline V_{,\gamma_1 \dots \gamma_n} \, \phi^{\gamma_1} \ldots \phi^{\gamma_n} \,.
\label{eq:GenLagExpanded}
\end{equation}
We further take $\overline V_{,\alpha\beta}$ and $\overline g_{\alpha\beta}$ to be diagonal. Their ratio then gives the masses of the fields
\begin{equation}
\overline V_{,\alpha\beta} = \overline{g}_{\alpha\beta} m_\alpha^2 \,.
\end{equation}
Then~\cref{eq:GenLagExpanded} yields the momentum space Feynman rules
\begin{subequations}
\begin{align}
\mathord{\begin{tikzpicture}[baseline=-0.65ex]
\draw (0,0) node[right] {$\beta$} -- (-1,0) node[left] {$\alpha$};
\end{tikzpicture}}
&= \frac{i \overline g^{\alpha\beta}}{p^2-m_\alpha^2} \,,
\label{eq:FeynmanPropagator}
\\[20pt]
\mathord{\begin{tikzpicture}[baseline=-0.65ex]
    \foreach \ang/\ind in {0/{$1,\alpha_1$},50/{$2,\alpha_2$},310/{$n,\alpha_n$}} {
\draw (0,0) -- (\ang:1);
\draw (\ang:1) node[inner sep=0pt,label={[label distance=0cm]\ang:\ind}] {};
\draw[dotted,thick] (70:0.8) arc (70:290:0.8);
};
\end{tikzpicture}}
&= -i \overline V_{, \alpha_1 \dots \alpha_n } - i \sum_{1 \leq i < j \leq n} p_{i} \cdot p_{j} \, \overline g_{\alpha_i \alpha_j, \alpha_1 \dots \hat \alpha_i \dots \hat \alpha_j \dots \alpha_n}\nonumber\\
&=  -i \overline V_{, \alpha_1 \dots \alpha_n }  -i  \sum_{1 \leq i < j \leq n} \left( \frac12 s_{ij} - \frac12 p_i^2 -\frac12 p_j^2 \right) \, \overline g_{\alpha_i \alpha_j, \alpha_1 \dots \hat \alpha_i \dots \hat \alpha_j \dots \alpha_n}  \nonumber \\[20pt]
&= -i  \overline V_{, \alpha_1 \dots \alpha_n }  -i  \sum_{1 \leq i < j \leq n} \frac12 s_{ij} \, \overline g_{\alpha_i \alpha_j, \alpha_1 \dots \hat \alpha_i \dots \hat \alpha_j \dots \alpha_n} \nonumber \\[15pt]
&\hspace{69pt}+i \sum_{1 \leq i \leq n} (n-1) m_i^2 \, \overline g_{\alpha_i (\alpha_1, \, \dots \hat \alpha_i \dots \alpha_n)} \nonumber\\[15pt]
&\hspace{69pt}+i \sum_{1 \leq i \leq n} (n-1) \left( p_i^2 - m_i^2 \right) \, \overline g_{\alpha_i (\alpha_1, \, \dots \hat \alpha_i \dots \alpha_n)} \,\,,
\label{eq:FeynmanVertex}
\end{align}
\end{subequations}
where all momenta are taken to be ingoing, and $\alpha_1 \dots \hat \alpha_i \dots \hat \alpha_j \dots \alpha_n$ should be interpreted as the string $\alpha_1 \dots \alpha_n$ with the hatted $\alpha_i,\alpha_j$ omitted.

The $S$-matrix is invariant under field redefinitions of the type in \cref{eq:CoordRedef}, \ie, coordinate changes.
This is realized at the level of individual amplitudes: they are covariant under coordinate changes, up to the external wavefunction renormalization factors that depend non-trivially on $\overline g_{\alpha\beta}$.
Namely, amplitudes assemble into products of kinematic factors, and covariant derivatives\footnote{The covariant derivative is defined with respect to the metric connection.  There are no gauge interactions in the models studied here.} of the potential and Riemann curvature tensor. To see this manifest covariance, we can analyze the simplest two examples: three-point and four-point amplitudes.

\vspace{10pt}
\noindent \textbf{Three-point amplitude:}
The three point amplitude is given by~\cite{Nagai:2019tgi}\footnote{We use a convention where $-i \amp = \text{residue factors} \times \text{Feynman rules}$, the Feynman rules being defined in \cref{eq:FeynmanPropagator,eq:FeynmanVertex}. This yields an extra minus sign as compared to typical conventions.}
\begin{align}
  \left( \prod_{\mu=\alpha,\beta\gamma} \overline g_{\mu\mu}^{\sfrac{1}{2}} \right) \amp_3 &=
\mathord{\begin{tikzpicture}[baseline=-0.65ex]
    \foreach \ang/\ind in {0/{$1,\alpha$},120/{$2,\beta$},240/{$3,\gamma$}} {
\draw (0,0) -- (\ang:1);
\draw (\ang:1) node[inner sep=0pt,label={[label distance=0cm]\ang:\ind}] {};
};
\node[circle,draw=black,thick,fill=white] at (0,0) {$\amp$};
\end{tikzpicture}}\notag\\[15pt]
&=  \overline V_{,\alpha\beta\gamma} + p_1 \cdot p_2 \, \overline g_{\alpha\beta,\gamma} + p_2 \cdot p_3 \, \overline g_{\beta\gamma,\alpha} + p_3 \cdot p_1 \, \overline g_{\gamma\alpha,\beta} \nonumber \\[15pt]
&=  \overline V_{,\alpha\beta\gamma} + \frac12 m_\alpha^2 \left( \overline g_{\beta\gamma,\alpha} - \overline g_{\alpha\beta,\gamma} - \overline g_{\gamma\alpha,\beta} \right) \nonumber \\[8pt]
&\hspace{41.5pt} + \frac12 m_\beta^2 \left( \overline g_{\gamma\alpha,\beta} - \overline g_{\beta\gamma,\alpha} - \overline g_{\alpha\beta,\gamma} \right) \nonumber \\[8pt]
&\hspace{41.5pt}+ \frac12 m_\gamma^2 \left( \overline g_{\alpha\beta,\gamma} - \overline g_{\gamma\alpha,\beta} - \overline g_{\beta\gamma,\alpha} \right) \nonumber\\[15pt]
&= \overline V_{,\alpha\beta\gamma} - m_\alpha^2 \overline \Gamma_{\alpha\beta\gamma} - m_\beta^2 \overline \Gamma_{\beta\gamma\alpha} - m_\gamma^2 \overline \Gamma_{\gamma\alpha\beta} \nonumber \\[15pt]
&= \overline V_{,\alpha\beta\gamma} - 3 \overline V_{,\rho(\alpha} \overline \Gamma^\rho_{\beta\gamma)} \nonumber \\[15pt]
&= \overline V_{;(\alpha\beta\gamma)}\, .
\label{eq:ThreePointAmplitude}
\end{align}
This assembles into a symmetric covariant derivative of the potential, once the kinematic and geometric identities
\begin{subequations}
\begin{align}
p_1 \cdot p_2 &= \frac12 \left( p_3^2 - p_1^2 - p_2^2 \right) \,, \\[8pt]
\Gamma_{\alpha\beta\gamma} &= \frac12\left( g_{\alpha\beta,\gamma} + g_{\gamma\alpha,\beta} - g_{\beta\gamma,\alpha} \right) \,, \label{eq:lowerGammaDefn}
\end{align}
\end{subequations}
and their permutations have been utilized.

\vspace{10pt}
\noindent \textbf{Four-point amplitude:}
This same strategy applies to higher point amplitudes as well, although the results are of course more complicated.  For concreteness, we provide the general 4-point amplitude, which is given by \cite{Nagai:2019tgi}
\begin{flalign}
  \left( \prod_{i=1}^4 \overline g_{\alpha_i\alpha_i}^{\sfrac{1}{2}} \right) \amp_4 &=
\mathord{\begin{tikzpicture}[baseline=-0.65ex]
    \foreach \ang/\ind in {45/{$1,\alpha_1$},135/{$2,\alpha_2$},225/{$3,\alpha_3$},315/{$4,\alpha_4$}} {
\draw (0,0) -- (\ang:1);
\draw (\ang:1) node[inner sep=0pt,label={[label distance=0cm]\ang:\ind}] {};
};
\node[circle,draw=black,thick,fill=white] at (0,0) {$\amp$};
\end{tikzpicture}}\notag\\[10pt]
&= \overline{V}_{,\alpha_1\alpha_2\alpha_3\alpha_4} + \left[ \frac14 p_1\cdot p_2\; \overline{g}_{\alpha_1\alpha_2,\alpha_3\alpha_4} + \text{perms}(1234) \right]
&\notag\\[10pt]
&\quad + \Bigg\{ \bigg[ \overline{V}_{;\left(\alpha_1\alpha_2\alpha_5\right)} - \left(s_{12}-m_{\alpha_5}^2\right) \overline{g}_{\alpha_5\rho}\overline\Gamma_{\alpha_1\alpha_2}^{\rho} \bigg] \frac{\overline{g}^{\alpha_5\alpha_6}}{s_{12} - m_{\alpha_5}^2}
&\notag\\[5pt]
&\hspace{35pt}
\times \bigg[ \overline{V}_{;\left(\alpha_3\alpha_4\alpha_6\right)} - \left(s_{12}-m_{\alpha_6}^2\right) \overline{g}_{\alpha_6\lambda}\overline\Gamma_{\alpha_3\alpha_4}^{\lambda} \bigg] + \text{cycs}\left(234\right) \Bigg\} \,,&\label{eqn:4ptAmplitudeStart}
\end{flalign}
where ``perms'' (``cycs'') denotes (cyclic) permutations. The terms in the second line are from the contact contribution, and the terms inside the curly braces result from the sum over the three channels (captured by the cycs(234)) that result from connecting two 3-point vertices with one propagator. The $\alpha_5$ and $\alpha_6$ species indices correspond to the propagator and thus are summed over, and we used the (off-shell) 3-point vertex
\begin{align}
\mathord{\begin{tikzpicture}[baseline=-0.65ex]
    \foreach \ang/\ind in {0/{$1,\alpha_1$},120/{$2,\alpha_2$},240/{$3,\alpha_3$}} {
\draw (0,0) -- (\ang:1);
\draw (\ang:1) node[inner sep=0pt,label={[label distance=0cm]\ang:\ind}] {};
};
\node[circle,draw=black,thick,fill=white] at (0,0) {$\amp$};
\end{tikzpicture}}
= \overline{V}_{;\left(\alpha_1\alpha_2\alpha_3\right)} - \left[ \left(p_1^2-m_{\alpha_1}^2\right) \overline{g}_{\alpha_1\lambda}\overline\Gamma_{\alpha_2\alpha_3}^{\lambda} + \text{cycs}\left(123\right) \right] \,,
\end{align}
see~\cref{eq:FeynmanVertex} above. We can simplify this expression by organizing the terms in~\cref{eqn:4ptAmplitudeStart}:
\begin{flalign}
\left( \prod_{i=1}^4 \overline g_{\alpha_i\alpha_i}^{\sfrac{1}{2}} \right) \amp_4
&=\frac{1}{24} \Bigg[
\overline{V}_{,\alpha_1\alpha_2\alpha_3\alpha_4}
- 6 \overline\Gamma_{\alpha_1\alpha_2}^\rho \overline{V}_{;\left(\rho\alpha_3\alpha_4\right)}
+ 3 \left(s_{12}-m_{\alpha_1}^2-m_{\alpha_2}^2\right) \overline{g}_{\alpha_1\alpha_2,\alpha_3\alpha_4}
&\notag\\[8pt]
&\hspace{40pt}
+ 3 \left(s_{12}-m_{\alpha_5}^2\right) \overline{g}_{\alpha_5\rho} \overline\Gamma_{\alpha_1\alpha_2}^\rho \overline\Gamma_{\alpha_3\alpha_4}^{\alpha_5}
&\notag\\[8pt]
&\hspace{40pt}
+ 3 \overline{V}_{;\left(\alpha_1\alpha_2\alpha_5\right)} \frac{\overline{g}^{\alpha_5\alpha_6}}{s_{12} - m_{\alpha_5}^2} \overline{V}_{;\left(\alpha_3\alpha_4\alpha_6\right)}
+ \text{perms}(1234) \Bigg]
&\notag\\[10pt]
&\hspace{0pt}
= \frac{1}{24} \Bigg[
\overline{V}_{,\alpha_1\alpha_2\alpha_3\alpha_4}
- 6 \overline\Gamma_{\alpha_1\alpha_2}^\rho \overline{V}_{;\left(\rho\alpha_3\alpha_4\right)}
- 3 \overline\Gamma_{\alpha_1\alpha_2}^\rho \overline\Gamma_{\alpha_3\alpha_4}^\lambda \overline{V}_{,\rho\lambda}
&\notag\\[8pt]
&\hspace{40pt}
- 6 m_{\alpha_1}^2 \overline{g}_{\alpha_1\alpha_2,\alpha_3\alpha_4}
+ 3 s_{12} \left( \overline{g}_{\alpha_1\alpha_2,\alpha_3\alpha_4} + \overline{g}_{\rho\lambda} \overline\Gamma_{\alpha_1\alpha_2}^\rho \overline\Gamma_{\alpha_3\alpha_4}^\lambda \right)
&\notag\\[8pt]
&\hspace{40pt}
+ 3 \overline{V}_{;\left(\alpha_1\alpha_2\alpha_5\right)} \frac{\overline{g}^{\alpha_5\alpha_6}}{s_{12} - m_{\alpha_5}^2} \overline{V}_{;\left(\alpha_3\alpha_4\alpha_6\right)}
+ \text{perms}(1234)
\Bigg] \,. &
\end{flalign}
To get the second line above, we have separated the $s_{ij}$ dependence from the mass dependence and have utilized $m_\alpha^2 \overline{g}_{\alpha\beta} = \overline{V}_{,\alpha\beta}$. Finally, we can use the relations
\begin{subequations}
\begin{flalign}
24 \overline{V}_{;\left(\alpha_1\alpha_2\alpha_3\alpha_4\right)} &=
\overline{V}_{,\alpha_1\alpha_2\alpha_3\alpha_4}
- 6 \overline\Gamma_{\alpha_1\alpha_2}^\rho \overline{V}_{;\left(\rho\alpha_3\alpha_4\right)}
- 3 \overline\Gamma_{\alpha_1\alpha_2}^\rho \overline\Gamma_{\alpha_3\alpha_4}^\lambda \overline{V}_{,\rho\lambda}
&\notag\\[5pt]
&\hspace{15pt}
- 4 m_{\alpha_1}^2 \overline{g}_{\alpha_1\alpha_2,\alpha_3\alpha_4}
+ 2 m_{\alpha_3}^2 \overline{g}_{\alpha_1\alpha_2,\alpha_3\alpha_4}
&\notag\\[5pt]
&\hspace{15pt}
+ 4 m_{\alpha_1}^2 \overline{g}_{\rho\lambda} \overline\Gamma_{\alpha_1\alpha_2}^\rho \overline\Gamma_{\alpha_3\alpha_4}^\lambda
+ \text{perms}(1234)
\,,&\\[10pt]
2 s_{12} \overline{R}_{\alpha_1\left(\alpha_3\alpha_4\right)\alpha_2} + \text{perms}(1234) &=
3 s_{12} \left( \overline{g}_{\alpha_1\alpha_2,\alpha_3\alpha_4} + \overline{g}_{\rho\lambda} \overline\Gamma_{\alpha_1\alpha_2}^\rho \overline\Gamma_{\alpha_3\alpha_4}^\lambda \right)
&\notag\\[5pt]
&\hspace{15pt}
- 2 \left( m_{\alpha_1}^2 + m_{\alpha_3}^2 \right) \overline{g}_{\alpha_1\alpha_2,\alpha_3\alpha_4}
&\notag\\[5pt]
&\hspace{15pt}
- 4 m_{\alpha_1}^2 \overline{g}_{\rho\lambda} \overline\Gamma_{\alpha_1\alpha_2}^\rho \overline\Gamma_{\alpha_3\alpha_4}^\lambda
+ \text{perms}(1234)
\,,&
\end{flalign}
\end{subequations}
to arrive at
\begin{flalign}
\left( \prod_{i=1}^4 \overline g_{\alpha_i\alpha_i}^{\sfrac{1}{2}} \right) \amp_4
=& \frac{1}{24} \Bigg[
\overline{V}_{;\left(\alpha_1\alpha_2\alpha_3\alpha_4\right)}
+ 2 s_{12} \overline{R}_{\alpha_1\left(\alpha_3\alpha_4\right)\alpha_2}
\notag\\[3pt]
&\hspace{40pt}
+ 3 \overline{V}_{;\left(\alpha_1\alpha_2\alpha_5\right)} \frac{\overline{g}^{\alpha_5\alpha_6}}{s_{12} - m_{\alpha_5}^2} \overline{V}_{;\left(\alpha_3\alpha_4\alpha_6\right)}
+ \text{perms}(1234) \Bigg]
&\notag\\[8pt]
=& \hspace{3pt} \overline{V}_{;\left(\alpha_1\alpha_2\alpha_3\alpha_4\right)}
+ \frac23 \left(
  s_{12} \overline{R}_{\alpha_1\left(\alpha_3\alpha_4\right)\alpha_2} +
  s_{13} \overline{R}_{\alpha_1\left(\alpha_2\alpha_4\right)\alpha_3} +
  s_{14} \overline{R}_{\alpha_1\left(\alpha_2\alpha_3\right)\alpha_4}
\right)  \notag\\[3pt]
&+ \text{factorizable pieces}
\,. &
\end{flalign}
Again, we see that the amplitude can be expressed entirely in terms of covariant derivatives of the potential and Riemann tensor, and propagator factors.  In particular, no explicit factors of the Christoffel symbols $\Gamma$ appear.  As we will see next, this is a general feature of the tree-level amplitudes, and is due to the fact that we can always go to the basis specified by normal coordinates (see \cref{app:NormalCoordinates}).

\pagebreak
\noindent \textbf{$n$-point amplitude:}
To construct the general $n$-point amplitude, the brute force approach to simplifying the calculation we used for the two examples above is untenable.  Therefore, we will leverage our ability to transform the Lagrangian to a specific basis, a field redefinition to ``normal coordinates,'' since partial derivatives of the potential and metric take a simpler form in terms of covariant quantities. In normal coordinates, the following are true
\begin{subequations}
\begin{align}
\overline g_{\alpha (\beta_1, \, \dots \beta_n) } \vert_{\substack{\text{normal} \\ \text{coords}}} &= 0 \,,
\label{eq:NoOffShellDeformation} \\[10pt]
\overline V_{,\gamma_1 \dots \gamma_n} \vert_{\substack{\text{normal} \\ \text{coords}}} &= \overline V_{;(\gamma_1 \dots \gamma_n)} \,, \label{eq:CovariantPotential} \\[8pt]
\overline g_{\alpha\beta,\gamma_1 \dots \gamma_n} \vert_{\substack{\text{normal} \\ \text{coords}}} &= 2\, \frac{n-1}{n+1}\, \overline{R}_{\alpha (\gamma_1 \gamma_2| \beta;| \gamma_3 \dots \gamma_n)} + \mathcal{O} \big( \s\overline{R}{\s}^{2} \big) \,, \label{eq:CovariantMetric}
\end{align}
\end{subequations}
where indices inside the vertical bars ``$|\cdots|$'' are excluded from the symmetrization. We refer the reader to \cref{app:NormalCoordinates} for derivation details.

\cref{eq:NoOffShellDeformation} implies that the last two parts of the Feynman vertices in \cref{eq:FeynmanVertex} vanish in normal coordinates. This means vertices are invariant under momentum deformations that take the external legs off-shell, but leave the Mandelstam invariants unchanged. Each Feynman graph built from these normal coordinate vertices will therefore have distinct kinematics, with kinematic poles corresponding to each propagator denominator; for contrast, see the cancellation of kinematic poles when using generic coordinates in \cref{eqn:4ptAmplitudeStart}.

Using \cref{eq:CovariantPotential,eq:CovariantMetric} for the remaining vertex terms then allows us to write
\vspace{4pt}
\begin{tcolorbox}[colback=white, colframe=black]
\begin{align}
  \hspace{-5pt}  \left( \prod_{i=1}^n \overline g_{\alpha_i \alpha_i}^{\sfrac 12} \right) \amp_n =&\, \overline V_{;(\alpha_1 \dots \alpha_n)} + \sum_{1\leq i < j \leq n} s_{ij} \left( \tfrac{n-3}{n-1} \right) \Big[ \overline R_{\alpha_i (\alpha_1 \alpha_2 | \alpha_j; |\alpha_3 \dots \hat \alpha_i \dots \hat \alpha_j \dots \alpha_n)} + \mathcal{O}\big( \s\overline{R}{\s}^{2} \big) \Big]  \nonumber \\[4pt]
  &\hspace{46pt}\,+ \text{factorizable pieces}\,,
  \label{eq:MasterGeometricAmp}%
\end{align}%
\end{tcolorbox}%
\noindent which, being a manifestly covariant amplitude, is true in any basis. Kinematically, the factorizable pieces are rational functions of Mandelstam invariants with non-trivial denominators; they have geometric coefficients built out of contractions of covariant derivatives of $V$ and the Riemann tensor.

We emphasize that the wavefunction renormalization factors on the LHS of \cref{eq:MasterGeometricAmp} appear to account for the diagonal but non-canonical kinetic term. The repeated indices on the LHS are \emph{not} summed. The product of the amplitude and these factors results in the manifestly covariant RHS.

\section{Geometrizing HEFT}
\label{sec:GeoHEFT}

Now that we have explored the general approach to expressing amplitudes in terms of geometric quantities, we will turn to our main application.  Specifically, from here forward we will study the EFT that describes the scalar sector of the Standard Model in the broken phase, the Higgs Effective Field Theory (HEFT), under the simplifying assumption of exact custodial symmetry.  In this section, we review the formulation of HEFT and explicitly construct some of its amplitudes using~\cref{eq:MasterGeometricAmp}.

\subsection{HEFT Manifold}

The custodially symmetric HEFT Lagrangian is a co-set construction $O(4)/O(3)$ relevant for a custodially symmetric model of electroweak symmetry breaking.
It can be written in terms of coordinates
\begin{align}
\vec \phi = \left( h , \pi_1 , \pi_2 , \pi_3 \right) \,,
\end{align}
where $h$ is the Higgs boson and $\vec{\pi}$ are the three Goldstone modes~\cite{Alonso:2015fsp,Alonso:2016oah}.  The most general HEFT Lagrangian including up to two derivatives is given by
\begin{equation}
\lag = \frac12 K(h)^2 \left( \partial h\right)^2 + \frac12 \left( v F(h) \right)^2 \left( \partial \vec n\right)^2 - V(h) + \mathcal{O}\big( \partial^4 \big) \,,
\label{eq:HEFTLagrangian}
\end{equation}
where $\vec{n}$ is the dimensionless non-linearly constrained field that contains the Goldstones:
\begin{equation}
  v \, \vec n = \begin{pmatrix}
    \pi_1 \\
    \pi_2 \\
    \pi_3 \\
    \sqrt{v^2 - \vec \pi \cdot \vec \pi}
  \end{pmatrix} \, , \qquad \text{where} \qquad \vec\pi\cdot\vec\pi = \sum_{i=1}^3 \left(\pi_i \right)^2 \,.
\end{equation}
The Lagrangian has the most general form compatible with invariance under the custodial symmetry transformation
\begin{equation}
  h \to h \,, \qquad \text{and} \qquad \vec n \to O \cdot \vec n\, ,
\end{equation}
where $O \in O(4)$.
The components of the metric as determined by \cref{eq:HEFTLagrangian} are
\begin{subequations}
\begin{align}
g_{hh} &= \left( K(h) \right)^2 \,, \\[3pt]
g_{\pi_i \pi_j} &= \left( F(h) \right)^2 \left[ \delta_{ij} + \frac{\pi_i \pi_j}{v^2 - \vec \pi \cdot \vec \pi} \right] \,, \\[3pt]
g_{h \pi_i} &= 0 \,.
\end{align}
\label{eq:MetricComponents}%
\end{subequations}%
Here and throughout, we use Latin indices $i,j,k,\ldots \in \{1,2,3\}$ to distinguish the three fields $\pi_i$. We can then use~\cref{eq:MetricComponents} to derive the metric connection
\begin{subequations}
\begin{align}
\Gamma^h_{hh} &= \frac{K^\prime}{K} \,, \\[4pt]
\Gamma^h_{\pi_i\pi_j} &= -\frac{F^\prime}{F K^2} g_{\pi_i\pi_j} \,, \\[4pt]
\Gamma^{\pi_i}_{\pi_j h} &= \Gamma^{\pi_i}_{h\pi_j} = \frac{F^\prime}{F} \delta^i_j \,, \\[4pt]
\Gamma^{\pi_i}_{\pi_j \pi_k} &= \frac{\pi_i}{\left(v F\right)^2}  g_{\pi_j \pi_k} \,.
\end{align}
\label{eq:ConnectionComponents}%
\end{subequations}%
Additionally, we will sometimes avail ourselves of the freedom to canonically normalize the Higgs coordinate via the field redefinition
\begin{equation}
  \canonh = Q(h) = \int^h_0 \dd \tilde{h} \, K(\tilde{h}) \,,
  \label{eq:RealCanonhDefinition}
\end{equation}
resulting in the Lagrangian
\begin{equation}
  \lag = \frac12 \left( \partial \canonh \right)^2 + \frac12 \Big( v \tilde F(\canonh) \Big)^2 \left( \partial \vec n\right)^2 - \tilde V(\canonh) + \mathcal{O}\big( \partial^4 \big) \,.
\end{equation}
General formulae may be reduced to this specific case via the substitution
\begin{align}
\big\{h,K,F,V\big\} \quad \mapsto \quad \big\{\canonh,1,\tilde F,\tilde V\big\} \,,
\end{align}
where $\tilde F = F \circ Q^{-1}$ and $\tilde V = V \circ Q^{-1}$.

Owing to the $O(4)$ symmetry, the HEFT manifold\footnote{The HEFT manifold is locally the product space of a line segment, parameterized by $h$, and a 3-sphere, parameterized by $\pi_i$~\cite{Alonso:2016oah}.} has only two independent sectional curvatures, $\mathcal{K}_{h}$ and $\mathcal{K}_{\pi}$.  These are related to the components of the Riemann curvature tensor by
\begin{subequations}
\begin{align}
  R_{\pi_i h h \pi_j} &= - g_{hh} g_{\pi_i \pi_j} \, \mathcal{K}_h \,, \\[3pt]
  R_{\pi_i \pi_k \pi_l \pi_j} &= \big( g_{\pi_i \pi_l} g_{\pi_k \pi_j} - g_{\pi_i \pi_j} g_{\pi_k \pi_l} \big) \, \mathcal{K}_\pi \,,
\label{eq:CurvatureComponents}
\end{align}
\end{subequations}
where the sectional curvatures are functions of the HEFT form factors:
\begin{subequations}\label{eq:SecCurv}
\begin{align}
\mathcal{K}_h& = -\frac{1}{K^2} \left[ \frac{F^{\prime\prime}}{F} - \frac{K^\prime}{K} \frac{F^\prime}{F} \right] \,, \\[6pt]
\mathcal{K}_\pi  &= \frac{1}{\left(vF\right)^2} \left[ 1 - \frac{\left(vF^\prime\right)^2}{K^2} \right] \,.
\end{align}
\end{subequations}
The Ricci scalar can then be determined by the sectional curvatures:
\begin{equation}
  R = 6\s \big(\mathcal{K}_h + \mathcal{K}_\pi\big) \, .
\end{equation}
The Laplacian of a scalar quantity is given by
\begin{equation}
    \nabla^2 V = \left( \frac{1}{K} \partial_h \right)^2 V + 3 \frac{F^\prime}{F K} \left( \frac{1}{K} \partial_h \right)  V \, .
\end{equation}
This provides the relevant formalism to characterize the geometry of the HEFT manifold. In the next section, we will use these relations to write the general geometrized amplitudes given above in~\cref{eq:MasterGeometricAmp} in terms of HEFT specific curvature invariants.

\subsection{HEFT Amplitudes\label{sec:HEFTAmplitudes}}

In~\cref{eq:MasterGeometricAmp} above, we derived an expression for the leading non-factorizable contribution to the tree-level amplitude for $n$-particle scattering that was expressed in terms of covariant derivatives of the potential and Riemann curvature tensor.  In this section, we will assume that we are working with HEFT, which will allow us to simplify these expressions to derive a useful form of $\amp( \pi_i \pi_j h^{n-2} )$ and $\amp( \pi_i \pi_j \pi_k \pi_l h^{n-4} )$ in terms of covariant derivatives of $V$ and the sectional curvatures $\mathcal{K}_h$ and $\mathcal{K}_\pi$.

To begin, we make the simple observation that if
\begin{equation}
  V_{; \beta_1 \dots \beta_n} \neq 0  \qquad \text{and} \qquad
  R_{\alpha_1 \alpha_2 \alpha_3 \alpha_4 ; \beta_1 \dots \beta_n}\neq 0 \, ,
\end{equation}
then an even number of the indices must be $\pi_i$ coordinates, as follows from the symmetries of \cref{eq:HEFTLagrangian}. Moreover, quantities with even numbers of $\pi$ indices are expressible as products of the $\pi$ metric and derivatives of $K$ and $F$ form factors.


Note also the symmetries of the curvature tensor:\footnote{These follow iteratively from the symmetries of the undifferentiated tensor, which are preserved under covariant differentiation, \eg
\begin{align*}
  R_{\alpha_1 \alpha_2 \alpha_3 \alpha_4; \beta} &= R_{\alpha_1 \alpha_2 \alpha_3 \alpha_4, \beta}
   -R_{\rho \alpha_2 \alpha_3 \alpha_4} \Gamma^\rho_{\beta \alpha_1}
  -R_{\alpha_1 \rho\alpha_3 \alpha_4} \Gamma^\rho_{\beta \alpha_2}
  -R_{\alpha_1 \alpha_2 \rho\alpha_4} \Gamma^\rho_{\beta \alpha_3}
  -R_{\alpha_1 \alpha_2 \alpha_3 \rho} \Gamma^\rho_{\beta \alpha_4} \,, \\
                                                 &= - \left( R_{\alpha_2 \alpha_1 \alpha_3 \alpha_4, \beta}
   -R_{\alpha_2 \rho \alpha_3 \alpha_4} \Gamma^\rho_{\beta \alpha_1}
  -R_{\rho \alpha_1 \alpha_3 \alpha_4} \Gamma^\rho_{\beta \alpha_2}
  -R_{\alpha_2 \alpha_1 \rho\alpha_4} \Gamma^\rho_{\beta \alpha_3}
  -R_{\alpha_2 \alpha_1 \alpha_3 \rho} \Gamma^\rho_{\beta \alpha_4} \right) \,, \\
                                                 &=R_{[\alpha_1 \alpha_2] \alpha_3 \alpha_4; \beta} \, .
\end{align*}
}
\begin{equation}
  R_{\alpha_1 \alpha_2 \alpha_3 \alpha_4 ; \beta_1 \dots \beta_n} \equiv  R_{[\alpha_1 \alpha_2] \alpha_3 \alpha_4 ; \beta_1 \dots \beta_n} \equiv  R_{\alpha_1 \alpha_2 [\alpha_3 \alpha_4] ; \beta_1 \dots \beta_n} \equiv  R_{\alpha_3 \alpha_4 \alpha_1 \alpha_2 ; \beta_1 \dots \beta_n} \, ,
\end{equation}
which imply, for instance, that at most two of the indices $\alpha_1, \alpha_2, \alpha_3, \alpha_4$ may be $h$ coordinates.
In what follows, we will use the above properties along with kinematic restrictions to simplify the general formula in~\cref{eq:MasterGeometricAmp}.

\subsubsection{Amplitude for 2 Goldstones and $n-2$ Higgses}

We begin with the $n$-point amplitude with 2 Goldstones and $n-2$ Higgses, $\amp\left( \pi_i \pi_j h^{n-2} \right)$.  We label the respective pion momenta by $p_1$ and $p_2$, and that of the Higgses as $p_3$ through $p_n$. Then~\cref{eq:MasterGeometricAmp} implies
\begin{align}
  \overline g_{\pi_i \pi_i}^{\sfrac{1}{2}} \overline g_{\pi_j \pi_j}^{\sfrac{1}{2}} \overline g_{hh}^{\frac{n-2}{2}}
  \amp\left( \pi_i \pi_j h^{n-2} \right) =&\, \overline V_{;(\pi_i \pi_j h^{n-2})} \nonumber \\[4pt]
  &+ \left({ \tfrac{n-3}{n-1} }\right) \bigg[ \overline R_{\pi_i (h h | \pi_j; | h^{n-4})} \, s_{12} + \overline R_{\pi_i (\pi_j h | h; | h^{n-4} )} \sum_{3 \leq k \leq n} s_{1k} \nonumber \\[6pt]
                                          &\hspace{20pt}+ \overline R_{\pi_j (\pi_i h | h; | h^{n-4} )} \sum_{3 \leq k \leq n} s_{2k} + \overline R_{h (\pi_i \pi_j | h; | h^{n-4} )} \sum_{3 \leq k < l \leq n} s_{kl} \bigg] \nonumber \\[4pt]
                                          &+ \mathcal{O} \big(\s\overline{R}{\s}^{2}\big) + \text{factorizable pieces} \,,
\end{align}
where all quantities are evaluated at the origin, and there is no sum on the repeated indices of the LHS.

Using the symmetries of the Riemann curvature tensor
\begin{subequations}
\begin{align}
  R_{\pi_i (h h | \pi_j; | h^{n-4})} &=  R_{\pi_i h h \pi_j; h^{n-4}} \,, \\[4pt]
  R_{\pi_i (\pi_j h | h; | h^{n-4} )} &= -\frac{1}{n-2} R_{\pi_i h h \pi_j; h^{n-4} } \,, \\[4pt]
  R_{h (\pi_i \pi_j | h; | h^{n-4} )} &=  \frac{2}{(n-2)(n-3)} R_{\pi_i h h \pi_j; h^{n-4} } \, ,
\end{align}
\end{subequations}
as well as the kinematic identities (noting that the pions are massless; the Higgses have mass $m_h$)
\begin{subequations}
\begin{align}
  \sum_{3 \leq k \leq n} \big( s_{1k} + s_{2k} \big)
   &= -2 s_{12} + 2(n-2) m_h^2  \,, \\[4pt]
  \sum_{3 \leq k < l \leq n} s_{kl}
    &= s_{12} + (n-2)(n-4) m_h^2 \,,
\end{align}
\end{subequations}
we find
\begin{align}
  \overline g_{\pi_i \pi_i}^{\sfrac{1}{2}} \overline g_{\pi_j \pi_j}^{\sfrac{1}{2}} \overline g_{hh}^\frac{n-2}{2}
  \amp\left( \pi_i \pi_j h^{n-2} \right) =& \overline V_{;(\pi_i \pi_j h^{n-2} )}
  + \overline R_{\pi_i h h  \pi_j;  h^{n-4} } \bigg( s_{12} - \frac{2 m_h^2}{n-1} \bigg) \nonumber \\[4pt]
                                          &+ \mathcal{O}  \big(\s\overline{R}{\s}^{2}\big) + \text{factorizable pieces} \, .
                                          \label{eq:TwoGoldstoneAmplitude}
\end{align}

We now construct an explicit expression for the $n$ index quantity $R_{\pi_i h h \pi_j; h^{n-4} }$ in terms of derivatives of the sectional curvature $\mathcal{K}_h$ taken with respect to the canonically normalized coordinate $\canonh$.  We begin by noting that $R_{\pi_i h h \pi_j; h^{n-4} }$ satisfies the recursion relation
\begin{align}
  \left( R_{\pi_i h h \pi_j; h^{n-4}} \right)_{;h} &= \left( \partial_h -(n-2)\Gamma^h_{hh} \right) R_{\pi_i h h \pi_j; h^{n-4}} - \Gamma^{\pi_k}_{\pi_i h} R_{\pi_k h h \pi_j; h^{n-4}} - \Gamma^{\pi_k}_{\pi_j h} R_{\pi_i h h \pi_k; h^{n-4}} \nonumber \\[6pt]
                                                 &= \bigg( \partial_h - 2 \frac{F^\prime}{F} - (n-2) \frac{K^\prime}{K} \bigg) R_{\pi_i h h \pi_j; h^{n-4}} \nonumber \\[6pt]
                                                 &= F^2 K^{n-2} \,\partial_h \bigg( \frac{R_{\pi_i h h \pi_j; h^{n-4}}}{F^2 K^{n-2}} \bigg) \nonumber \\[6pt]
                                                 &= F^2 K^{n-1} \,\partial_\canonh \bigg( \frac{R_{\pi_i h h \pi_j; h^{n-4}}}{F^2 K^{n-2}} \bigg) \, ,
\end{align}
where we have used the Christoffel symbols of \cref{eq:ConnectionComponents}, and the appearance of $\canonh$ in the last line relies on $\frac{1}{K}\partial_h = \partial_\canonh$. Repeated application of this recursion relation yields
\begin{equation}
  \frac{R_{\pi_i h h \pi_j; h^{n-4}}}{F^2 K^{n-2}} = \partial^{n-4}_\canonh \bigg( \frac{R_{\pi_i h h \pi_j}}{F^2 K^2} \bigg) = - \frac{g_{\pi_i\pi_j}}{F^2} \partial_\canonh^{n-4} \mathcal{K}_h \, .
\end{equation}
When evaluated at the origin, this reduces to
\begin{equation}
  \frac{\overline R_{\pi_i h h \pi_j; h^{n-4}}}{F(0)^2 K(0)^{n-2}} = - \delta_{ij} \, \partial_\canonh^{n-4} \mathcal{K}_h\big|_{\canonh=0} \,.
  \label{eq:NablaRiemannIsPartialSectional}
\end{equation}
To treat the potential piece, we use the fact that the commutator of covariant derivatives is proportional to the Riemann curvature tensor. This allows us to unsymmetrize the indices of the potential piece
\begin{equation}
  \overline V_{;(\pi_i \pi_j h^{n-2})} =
  \overline V_{;\pi_i \pi_j h^{n-2}} + \mathcal{O} \left( \overline{V} \overline{R} \right) \, ,
\end{equation}
at the expense of neglecting terms that are proportional to products of covariant derivatives of the potential and Riemann curvature tensors. An analogous recursion relation,
\begin{align}
  \left( V_{;\pi_i \pi_j h^{n-2}} \right)_{;h}     &= \bigg( \partial_h - 2 \frac{F^\prime}{F} - (n-2) \frac{K^\prime}{K} \bigg) V_{;\pi_i \pi_j h^{n-2}} \nonumber \\[6pt]
                                                 &= F^2 K^{n-2} \,\partial_h \bigg( \frac{V_{;\pi_i \pi_j h^{n-2}}}{F^2 K^{n-2}} \bigg)
                                                 = F^2 K^{n-1} \,\partial_\canonh \bigg( \frac{V_{;\pi_i \pi_j h^{n-2}}}{F^2 K^{n-2}} \bigg) \, ,
\end{align}
gives
\begin{equation}
  \frac{V_{;\pi_i \pi_j h^{n-2}}}{F^2 K^{n-2}} = \partial^{n-2}_\canonh \bigg( \frac{V_{;\pi_i \pi_j}}{F^2} \bigg) = \frac{g_{\pi_i\pi_j}}{ 3 F^2} \partial_\canonh^{n-2} \left( \nabla^2 V - \partial_\canonh^2 V \right) \, ,
\end{equation}
which reduces at the origin to
\begin{equation}
  \frac{\overline V_{; \pi_i \pi_j h^{n-2}}}{F(0)^2 K(0)^{n-2}} = \frac13 \, \delta_{ij} \, \partial_\canonh^{n-2} \left( \nabla^2 V - \partial_\canonh^2 V \right) \big|_{\canonh=0}\, .
  \label{eq:NablaPotentialIsPartialLaplacian}
\end{equation}

Dividing through by the wavefunction renormalization factors, the amplitude in \cref{eq:TwoGoldstoneAmplitude} therefore takes the form
\begin{align}
  \amp\left( \pi_i \pi_j h^{n-2} \right) &= \frac13\, \delta_{ij}\, \partial_\canonh^{n-2} \left( \nabla^2 V - \partial_\canonh^2 V \right) \big|_{\canonh=0}
  - \bigg( s_{12} - \frac{2 m_h^2}{n-1} \bigg) \delta_{ij}\, \left( \partial_\canonh^{n-4} \mathcal{K}_h \right) \big|_{\canonh=0} \notag\\[5pt]
                                          &\quad + \mathcal{O} \big(\s\overline{V}\s\overline{R}\s,\s\overline{R}{\s}^{2}\big) + \text{factorizable pieces} \,.
                                          \label{eq:TwoGoldstoneAmplitudeExplicit}
\end{align}
We see that the leading contribution to the amplitude with 2 Goldstones and $n-2$ Higgses in the high energy limit can be determined from successive partial derivatives of the potential $V$, its Laplacian $\nabla^2 V$, and the sectional curvature $\mathcal{K}_h$.  In the next section, we will leverage this result to determine the scale of unitarity violation.

\subsubsection{Amplitude for 4 Goldstones and $n-4$ Higgses}

Four-Goldstone amplitudes can be calculated in a similar way, albeit the details are a bit more tedious. In particular, neglecting pieces of the size $\mathcal{O} \big(\s\overline{V}\s\overline{R}\s,\s\overline{R}{\s}^{2}\big)$ again allows us to treat covariant derivatives as commuting objects when acting on potential and curvature components. With some further help of the second Bianchi identity (which one can also check using the explicit forms of $\mathcal{K}_h$ and $\mathcal{K}_\pi$ in \cref{eq:SecCurv}):
\begin{align}
& 3R_{\alpha_1\alpha_2\left[\alpha_3\alpha_4;\alpha_5\right]} = R_{\alpha_1\alpha_2\alpha_3\alpha_4;\alpha_5} + R_{\alpha_1\alpha_2\alpha_4\alpha_5;\alpha_3} + R_{\alpha_1\alpha_2\alpha_5\alpha_3;\alpha_4} = 0 \notag\\[8pt]
& \quad\Rightarrow\quad   2 \frac{F'}{F} \left(\mathcal{K}_h-\mathcal{K}_\pi\right) = \partial_h \mathcal{K}_\pi \,,
\end{align}
we can eventually write everything in terms of successive partial derivatives $\partial_\canonh$ acting on $V$, $\nabla^2 V$, $\nabla^4 V$, $\mathcal{K}_h$, $\nabla^2 \mathcal{K}_h$, and $\mathcal{K}_\pi$:
\begin{align}
\mathcal{A}\big(\pi_i \pi_j \pi_k \pi_l h^{n-4}\big) &=
\frac{1}{15} \left( \delta_{ij}\delta_{kl} + \delta_{ik}\delta_{jl} + \delta_{il}\delta_{jk} \right)
\partial_\canonh^{n-4} \left( \nabla^4 V - 2\partial_\canonh^2\nabla^2 V + \partial_\canonh^4 V \right) \big|_{\canonh=0}
\notag\\[8pt]
&\hspace{-60pt}
-\frac19 \left( \delta_{ij}\delta_{kl} + \delta_{ik}\delta_{jl} + \delta_{il}\delta_{jk} \right)
\left[ s_{1234} - \frac{12(n-4)m_h^2}{(n-1)(n-2)} \right]
\partial_\canonh^{n-6} \left( \nabla^2 \mathcal{K}_h - \partial_\canonh^2 \mathcal{K}_h \right) \big|_{\canonh=0}
\notag\\[8pt]
&\hspace{-60pt}
+ \bigg\{ \frac16 \left( \delta_{ij}\delta_{kl} + \delta_{ik}\delta_{jl} + \delta_{il}\delta_{jk} \right) s_{1234}
\notag\\[3pt]
&\hspace{-60pt}
\qquad - \frac12 \Big[ \delta_{ij}\delta_{kl} \left(s_{12}+s_{34}\right) + \delta_{ik}\delta_{jl} \left(s_{13}+s_{24}\right) + \delta_{il}\delta_{jk} \left(s_{14}+s_{23}\right) \Big] \bigg\}\,
\partial_\canonh^{n-4} \mathcal{K}_\pi \big|_{\canonh=0}
\notag\\[8pt]
&\hspace{-60pt}
+ \mathcal{O} \big(\s\overline{V}\s\overline{R}\s,\s\overline{R}{\s}^{2}\big) + \text{factorizable pieces} \,,
\label{eq:FourGoldstoneAmplitudeExplicit}
\end{align}
where $s_{ij \dots k} = (p_i + p_j + \ldots + p_k)^2$.
We see that this amplitude includes a new element as compared to \cref{eq:TwoGoldstoneAmplitudeExplicit}, in that it explicitly depends on (derivatives of) $\mathcal{K}_\pi$.  Furthermore, note that terms involving $\mathcal{K}_h$ in the second line only appear for $n\geq 6$.  Although we will not utilize this result explicitly in what follows, we will occasionally find it useful in our discussion of concrete models to note the role of $\mathcal{K}_\pi$ in the connection between the unitarity cutoff and the analyticity properties of the sectional curvatures.  A full analysis exploring the consequences of \cref{eq:FourGoldstoneAmplitudeExplicit} for the scale of unitarity violation is left for future work.

\section{Unitarity Violation From the Radius of Convergence}
\label{sec:UnitarityCutoffs}

In this Section, we will show that the radius of convergence $v_\star$ of the sectional curvature $\mathcal{K}_h$ can be translated into the scale of unitarity violation for SMEFT or HEFT. Although we will work in terms of the HEFT parameterization of \cref{eq:HEFTLagrangian}, our conclusions hold for both ``reducible HEFTs'' (which can be written as SMEFTs) and ``irreducible HEFTs'' (which cannot). Our argument leverages the Cauchy-Hadamard theorem, which provides a connection between the radius of convergence of a function and the growth of its derivatives.  Then using \cref{eq:TwoGoldstoneAmplitudeExplicit}, our formula that gives the HEFT amplitude for 2 Goldstone + $(n-2)$ Higgs scattering in terms of derivatives of $\mathcal{K}_h$ and $V$, we can derive a scale of unitarity violation in terms of $v_\star$.

This might seem like an overly mathematical reframing of standard results regarding unitarity violation of EFTs.  However, if we know the functional form of $\mathcal{K}_h$, the radius of convergence can also be determined by the location of the closest non-analyticity of the $\mathcal{K}_h$ function in the complex $\canonh$ plane. As we proved in~\cite{Cohen:2020xca}, the presence of such a non-analyticity sufficiently close to our vacuum tells us that we can not faithfully match a perturbative UV BSM theory onto SMEFT, and the larger HEFT parameterization is required, see \cref{sec:UVCompletions} below for examples.  This is the final ingredient that ties our story together: if we \emph{must} use HEFT, there will be a singularity on the scalar manifold at a distance $v_\star$ where $v_\star \sim v$, which in turn sets the scale of unitarity violation by way of the radius of convergence.  Through the examples of weakly coupled UV completions in \cref{sec:UVCompletions}, we will see that these non-analyticities of $\mathcal{K}_h$ are avatars of states which have been integrated out when constructing the EFT (\eg~the `Loryons' \cite{Banta:2021dek}), and which are responsible for restoring unitarity for the amplitudes as calculated by the UV theory.

The rest of this section is devoted to demonstrating how, in the limit that the number of $h$ indices is taken to be large, \cref{eq:NablaRiemannIsPartialSectional} can be used to relate the size of $R_{\pi_i h h \pi_j; h^{n-4}}$ to the radius of convergence of $\mathcal{K}_h$ in the complex plane of $\canonh$. This radius of convergence $v_\star$ sets the scale of perturbative unitarity violation by way of the amplitudes of \cref{sec:HEFTAmplitudes}, as we will show presently.

\subsection{Cauchy-Hadamard Theorem}

The Cauchy-Hadamard theorem links the convergence radius of a Taylor expansion with the growth rate of the expansion coefficients. More concretely, consider a single complex variable function $f(z)$. If it is analytic at a point $z_0$, it has a Taylor expansion
\begin{equation}
f(z) = \sum_{n=0}^\infty c_n (z-z_0)^n \,.
\end{equation}
The Cauchy-Hadamard theorem states that the convergence radius $z_\star$ of this Taylor expansion is given by the limit superior:
\begin{equation}
\frac{1}{z_\star} = \limsup_{n\to\infty} \left|c_n\right|^{1/n} \,.
\label{eqn:CauchyHadamard}
\end{equation}
A benchmark example is the following Taylor expansion at the origin
\begin{equation}
\frac{1}{1-az} = \sum_{n=0}^{\infty} a^n z^n \,,
\end{equation}
whose convergence radius should be $z_\star=1/a$ by \cref{eqn:CauchyHadamard}, agreeing with what we would infer from the location of the pole. Basically, the faster the expansion coefficients $c_n$ grow, the smaller the convergence radius will be, and the closer the non-analyticity will be to $z_0$. On the other hand, if a function has no non-analyticity across the whole complex plane, the Taylor expansion coefficient $c_n$ must fall faster than an exponential to make the RHS of \cref{eqn:CauchyHadamard} vanish. A simple example of this kind is
\begin{equation}
e^z = \sum_{n=0}^{\infty} \frac{1}{n!} z^n \,,
\end{equation}
where the expansion coefficients $c_n=1/n!$ fall factorially.

\subsection{Energy Growth of HEFT Amplitudes}
\label{sec:AmpGrowth}

Applying the Cauchy-Hadamard theorem to the function $\mathcal{K}_h(\canonh)$, we can link the limiting size of its derivatives in terms of its radius of convergence, $v_\star$ about the origin $\canonh=0$. Dividing through by $\overline{\mathcal{K}}_h \equiv \mathcal{K}_h|_{\canonh=0}$ to render the function dimensionless, we have
\begin{equation}
  \underset{n \to \infty}{\lim \sup} \left( \frac{1}{n!} \frac{ |\partial_\canonh^{n} \mathcal{K}_h| }{|\mathcal{K}_h|} \right)^{\sfrac{1}{n}} \bigg|_{\canonh=0} = \frac{1}{v_\star} \, .
\end{equation}
In light of the above, we introduce the quantity $a_n$:
\begin{equation}
\left( \frac{1}{n!} \frac{ |\partial_\canonh^n \mathcal{K}_h| }{|\mathcal{K}_h|} \right)^{\sfrac{1}{n}} \bigg|_{\canonh=0} = \frac{1}{a_n v_\star} \,,
\label{eq:LimitingScaleSectionalCurvature}
\end{equation}
which satisfies\footnote{For our purposes, the presence of `sup' (superior) and `inf' (inferior) in the above simply means that the limits may only be an arbitrary good approximation for \emph{some} large $n$, not \emph{all} large $n$, above a certain number, and will not be consequential in deriving the unitarity bounds.}
\begin{equation}
  \underset{n \to \infty}{\lim \inf} \, a_n = 1, \quad\text{ but generally }\quad
  \underset{n \to \infty}{\lim \inf} \, (a_n)^n \neq 1 \,.
  \label{eq:CauchyHadamardInTermsOfan}
\end{equation}
In practice, $a_n$ is sufficiently close to 1 for $n$ of `a few'; see the examples of \cref{sec:UVCompletions} and the plots of $a_n$ given in \cref{fig:an}.

\cref{eq:LimitingScaleSectionalCurvature} is redolent of power counting schemes in HEFT \cite{Gavela:2016bzc}, wherein adding a Higgs leg to an amplitude adds a factor $1/v_\star$. We stress here that $v_\star$ is not the same as $v$, the quantity which fixes $G_F$ once the remaining SM field content is incorporated into the Lagrangian \cref{eq:HEFTLagrangian}. This will be apparent from the examples of \cref{sec:UVCompletions}.

Recent works \cite{Chang:2019vez,Falkowski:2019tft,Abu-Ajamieh:2020yqi} have shown how BSM deviations from the Standard Model that are poorly described by a truncated SMEFT expansion must have perturbative unitarity cutoffs no higher than the TeV scale; in particular, \cite{Chang:2019vez,Falkowski:2019tft} link the failure of the SMEFT expansion to the presence of non-analyticities in the Lagrangian at $H=0$, which in turn cause inevitable unitarity violation at the TeV scale.

Below, we reprise these arguments by applying generic perturbative unitarity limits to the amplitude written in \cref{eq:TwoGoldstoneAmplitudeExplicit}, making use of the geometric formulation of amplitudes to write the steps in terms of expressly field redefinition invariant quantities. Combined with knowledge of the amplitudes' limiting size in \cref{eq:LimitingScaleSectionalCurvature}, we derive a unitarity cutoff at the scale $4 \pi v_\star$ for \emph{any} BSM theory. In \cref{sec:UVCompletions}, we pair this result with a geometric classification of theories which may be poorly described by SMEFT, from our previous work \cite{Cohen:2020xca}. As we show by example, theories that are well described by SMEFT may have $v_\star \gg v$, whereas theories that are not must have $v_\star \sim v$.

\begin{figure}
  \centering
  \includegraphics[align=c,height=0.3\textheight]{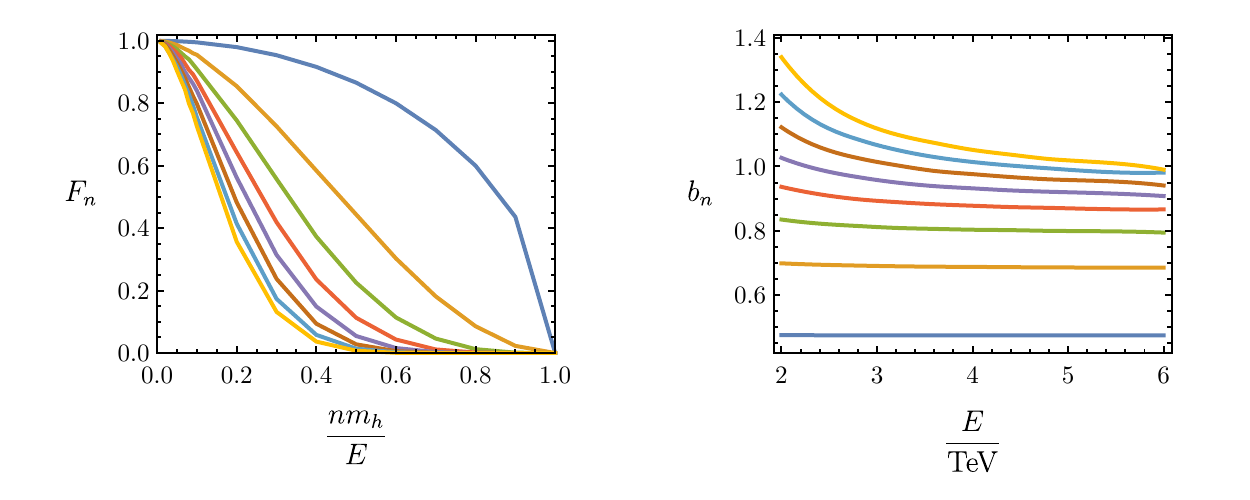}\\[10pt]
   $n= \includegraphics[align=c,width=0.8\textwidth]{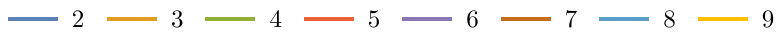}$
  \caption{Left: the ratio of the volume of $n$-body Higgs phase space (each with mass $m_h$) to $n$-massless-body phase space, where $E$ is the center-of-mass energy. The phase space volumes are calculated using a Cython implementation \cite{RAMBOCython} of the RAMBO algorithm \cite{Kleiss:1985gy}. Right: the derived factor $b_n(E)$, \cref{eq:bndef}, for TeV scale center-of-mass energies. Note that the $b_n$ factors are essentially flat in $E$ and order one, and thus do not parametrically alter the unitarity cutoffs \cref{eq:TwoGoldstoneUnitarityBound}.\label{fig:PhaseSpaceNumerics}}
\end{figure}

Following the insight of \cite{Falkowski:2019tft}, we now consider the unitarity cutoff resulting from the amplitude $\amp\left(\pi^2 \to h^n \right)$; however, in the interests of simplicity we focus on the effects of the two derivative terms, as opposed to the potential, whose effects we discuss later.
Ignoring the $\mathcal{O}\left(V\right)$ and $\mathcal{O}\left(R^2\right)$ terms, \cref{eq:TwoGoldstoneAmplitudeExplicit} has trivial momentum dependence, and can be rewritten in terms of a correctly normalized $s$-wave matrix element $\hat M$ using \cref{eq:swaveproj}:
\begin{align}
 1 > \big|\hat M_{\beta\alpha}\big|^2 &= \frac{\mathrm{Vol}_{2}}{2!} \frac{\mathrm{Vol}_n}{n!} | \amp(\pi_i \pi_j \to h^n ) |^2 \nonumber \\
                      &= \frac{1}{2 (8 \pi)^2 n! (n-1)! (n-2)!} \left( \frac{E}{4 \pi} \right)^{2(n-2)} F_n\left(\frac{n m_h^2}{E} \right) \nonumber \\
                      & \qquad \qquad \times E^4 \left( 1 - \frac{2 m_h^2}{(n+1) E^2} \right)^2
                      \left( \delta_{ij} |\partial_\canonh^{n-2} \mathcal{K}_h|^2 \right) \big|_{\canonh=0} \,,
  \label{eq:TwoGoldstoneMSquared}
\end{align}
where the inequality follows from the perturbative unitarity constraint given in \cref{eq:PerturbativeUnitarityIndividual}. Recall that $E$ is the center of mass energy. In \cref{eq:TwoGoldstoneMSquared}, we have also defined a factor $F_n \left( \frac{n m_h}{E} \right)$ to be the ratio of the phase space volume of $n$ Higgs particles (each having mass $m_h$) to the phase space volume of $n$ massless particles, see \cref{eq:NBodyMasslessPhaseSpace}. Its argument is normalized such that $F_n(0)=1$ and $F_n(1)=0$; in \cref{fig:PhaseSpaceNumerics} (left) we plot $F_n$ for $n = 2,\dots,9$.

Finally, we rearrange the matrix element to get
\begin{align}
1 > |\hat M_{\beta\alpha}|^2 &= \left(\frac{1}{a_{n-2}} \right)^{2(n-2)} \left(\frac{1}{b_n} \right)^{2n} \frac{1}{n!} \left( \frac{E}{4 \pi v_\star} \right)^{2n} \delta_{ij}\, \left|v_\star^2\,\overline{\mathcal{K}}_h\right|^2 \,,
\label{eq:FinalMSquared}
\end{align}
where $a_{n-2}$ encodes the scale of multiple derivatives of the sectional curvature and is defined in \cref{eq:LimitingScaleSectionalCurvature}, and $b_n$ absorbs the remaining phase space factors:
\begin{equation}
\left(\frac{1}{b_n} \right)^{2n} = \frac{(4\pi)^2}{8(n-1)} \left( 1 - \frac{2 m_h^2}{(n+1) E^2} \right)^2 F_n\left(\frac{n m_h^2}{E} \right) \, .
\label{eq:bndef}
\end{equation}
We plot $b_n(E)$ for $n= 2,\dots,9$ in \cref{fig:PhaseSpaceNumerics} (right).

Let us now consider the unitarity cutoff resulting from \cref{eq:FinalMSquared}. Setting $i=j$, with the use of Stirling's approximation to express the factorial as an approximate power (which is why Euler's constant $e$ appears), we place an upper bound on the center-of-mass energy at which the EFT amplitude is valid
\begin{equation}
  E < 4 \pi v_\star \times \left(a_{n-2}\right)^\frac{n-2}{n} \times b_{n} \times \left|v_\star^2\,\overline{\mathcal{K}}_h\right|^{-\sfrac{1}{n}} \times \sqrt{\frac{n}{e}} \,.
  \label{eq:TwoGoldstoneUnitarityBound}
\end{equation}
This is the central equation that we will use to compute the unitarity cutoff numerically for various examples. Let us summarize the key features of its factors one-by-one:
\begin{itemize}
\item The factor $4\pi v_\star$ serves as our intuitive estimate of the unitarity cutoff scale. We expect the other factors to be close to $1$ for $n$ of `a few'.
\item The factor $\left(a_{n-2}\right)^\frac{n-2}{n}$ stems from the quantity $a_{n-2}$ defined in \cref{eq:LimitingScaleSectionalCurvature}, which characterizes the growth of $\left( \partial_\canonh^{n-2} \mathcal{K}_h \right) \big|_{\canonh=0}$ involved in the two-Goldstone amplitude (see \cref{eq:TwoGoldstoneAmplitudeExplicit}). By the Cauchy-Hadamard theorem (see \cref{eq:CauchyHadamardInTermsOfan}), $a_{n-2}$ will tend to $1$ for large $n$. In practice, it is sufficiently close to $1$ for $n$ of `a few'; see the examples of \cref{sec:UVCompletions} and the plots of $a_n$ given in \cref{fig:an}.
\item The factor $b_n$ as defined in \cref{eq:bndef} is a collection of various kinematic factors. It is a very mild monotonically decreasing function of $E$, and \cref{fig:PhaseSpaceNumerics} (right) shows that in the TeV region of interest -- and for single digit $n$ -- it is approximately $1$ or less.
\item The factor $\left|v_\star^2\,\overline{\mathcal{K}}_h\right|^{-\sfrac{1}{n}}$ originates from the normalization factor $\left|v_\star^2\,\overline{\mathcal{K}}_h\right|$ for the curvature piece $\left( \partial_\canonh^{n-2} \mathcal{K}_h \right) \big|_{\canonh=0}$ involved in the amplitude. There are two scenarios for this factor. The ``weakly curved'' scenario happens when $\left|v_\star^2\,\overline{\mathcal{K}}_h\right| \ll 1$. In this case, the factor $\left|v_\star^2\,\overline{\mathcal{K}}_h\right|^{-\sfrac{1}{n}}$ reduces significantly as $n$ increases, motivating us to go to higher values of $n$ for a more stringent unitarity cutoff. We are mostly interested in this scenario. On the other hand, a sizable $\left|v_\star^2\,\overline{\mathcal{K}}_h\right| \sim 1$ signatures the ``strongly curved'' scenario. In this case, we could use the $n=2$ form of \cref{eq:TwoGoldstoneMSquared} to derive a unitarity cutoff of $E < 4 \pi \left|\overline{\mathcal{K}}_h\right|^{-\sfrac{1}{2}}$ from the undifferentiated sectional curvature \cite{Alonso:2015fsp}, which would be at least as stringent as the bounds resulting from higher values of $n$.
\end{itemize}
Putting all of this together, we have derived a unitarity bound of the following form:
\begin{flalign}
&&
\tcbhighmath[colback=white, colframe=black]{\begin{aligned}
E < 4 \pi v_\star \times \sqrt{\frac{n}{e}} \sim 4 \pi v_\star\,, \quad \text{for}\quad n \sim \mathcal{O}(\text{few})\,.
\label{eq:EUnitarityBound}
\end{aligned}}
&&
\end{flalign}
As demonstrated in \cite{Falkowski:2019tft}, we can further strengthen the bound, as well as reduce its dependence on the $ \left| v_\star^2\, \overline{\mathcal{K}}_h \right|$ factor, by summing over different final states per \cref{eq:PerturbativeUnitarity}. Retaining the approximation $a_n \sim b_n \sim 1$ in \cref{eq:FinalMSquared}, we get
\begin{flalign}
1 > \sum_n |\hat M_{\beta\alpha}|^2 &\sim \sum_n \frac{1}{n!} \left( \frac{E}{4 \pi v_\star} \right)^{2n} \delta_{ij} \left|v_\star^2\, \overline{\mathcal{K}}_h\right|^2 \sim \delta_{ij} \left|v_\star^2\, \overline{\mathcal{K}}_h\right|^2 \exp \left[ \left( \frac{E}{4 \pi v_\star} \right)^2 \right] .&
\label{eq:FRredux}
\end{flalign}
This implies that the unitarity cutoff $E \sim 4 \pi v_\star$ only depends logarithmically on the couplings and other dependences that determine the prefactor.

In the ``weakly curved'' regime $\left|v_\star^2\,\overline{\mathcal{K}}_h\right| \ll 1$, and $n \sim \mathcal{O}(\text{a few})$, let us consider the terms we are neglecting in \cref{eq:TwoGoldstoneMSquared}. One such class is higher order terms in $R$, arising from $\mathcal{O}(R^2)$ pieces in the amplitude \cref{eq:TwoGoldstoneAmplitudeExplicit}, either from the contact term or the factorizable pieces with non-trivial kinematics. Both of them appear with a factor
\begin{align}
  \amp(\pi_i\pi_j \to h^n ) &\supset \sum_m \frac{ \overline R_{\pi_i h h \pi_k; (h)^{n-4-m}} \, \overline g^{\pi_k \pi_l} \, \overline R_{\pi_l h h \pi_j; (h)^m} }{F^2 K^n} \times \text{kinematic factors} \nonumber \\
                            &= \sum_m \delta_{ij}\, \left(\partial_\canonh^{n-4-m} \mathcal{K}_h\right)\, \left(\partial_\canonh^{m} \mathcal{K}_h \right) \big|_{\canonh=0} \times \text{kinematic factors} \nonumber \\
                            &\sim \delta_{ij} \frac{\left(v_\star^2\,\overline{\mathcal{K}}_h\right)^2}{v_\star^n} \sum_m \frac{(n-4-m)!}{(a_{n-4-m})^{n-4-m}} \frac{m!}{(a_m)^m} \times \text{kinematic factors} \,,
\end{align}
where we used the limiting behaviour of derivatives of the sectional curvature given in \cref{eq:LimitingScaleSectionalCurvature}. Parametrically, the above is suppressed by $\left|v_\star^2\,\overline{\mathcal{K}}_h\right| < 1$ relative to the $\mathcal{O}(R)$ piece in the amplitude. However, the $\mathcal{O}(R^2)$ terms are combinatorially more numerous than the $\mathcal{O}(R)$ piece; the value of $n$ above which they numerically dominate bounds the validity of our perturbative calculations.\footnote{Using \cref{eqn:gNormalCoords}, the $\mathcal{O}(R^2)$ contribution from the contact terms can be calculated exactly:
\begin{equation}
  \amp(\pi_i\pi_j \to h^n) \supset \sum_{k=2}^{n-2} {n \choose k} \frac{k-1}{k+1} \frac{n-k-1}{n-k+1} \frac{(n-k)^2+k^2+n(n+2)}{2n(n-1)}\, s\, \delta_{ij}\, \left( \partial_\canonh^{n-4-m} \mathcal{K}_h \right)\, \left( \partial_\canonh^{m} \mathcal{K}_h \right) \big|_{\canonh=0} \,.
\end{equation}
We have checked numerically for $n \leq 9$ that, when projected into an $s$-wave state, the above terms are all larger than their counterpart $\mathcal{O}(R^2)$ term arising from factorizable tree graphs, namely those with Goldstones in the $t$-channel.}
By contrast, in the ``strongly curved'' regime $\left|v_\star^2\,\overline{\mathcal{K}}_h\right| \sim 1$, the $R^{n+1}$ are not negligible and instead have the same parametric size as the $\partial^{2n} R$ term considered in the bulk of this work.

We generically expect from studying weakly coupled UV completions (see the explicit examples of \cref{sec:UVCompletions}) that the potential terms in the amplitude have non-analyticities at the same points in the complex plane of $\canonh$ as the curvature terms, and therefore have the same radius of convergence $v_\star$ about our vacuum. This is because the resulting zero- and two-derivative terms in the EFT are simple functions of the Higgs-dependent mass $m^2(h)$ of the states that are integrated out, and these functions have singularities exactly when $m^2(h)=0$ \cite{Cohen:2020xca}.\footnote{This property should extend to higher derivative terms in the EFT that we do not consider in this work.} Moreover, the zero- and two-derivative terms have different parametric dependence on the couplings of any UV completion, and their linear combination in the $n$-Higgs amplitude should not cancel the factorial growth, $n!/v_\star^n$, exhibited by each of the terms individually. Although it would mildly change the energy dependence of the RHS of \cref{eq:FinalMSquared}, the inclusion of the potential terms should not significantly change the energy cutoff bound presented in~\cref{eq:EUnitarityBound}.

As always, we remain agnostic about the effects on the amplitude of the four-and-higher derivative terms in the Lagrangian. In principle, they may be encoded as four-and-higher index tensors on the field-space manifold, whose normal coordinate expansions will map onto $\mathcal{O}(s^2)$ and higher dependent contact terms in the amplitudes. From our explorations of perturbative matching examples, we also expect these tensors to have singularities at the same points in field space as the potential and metric. A full understanding of these terms and their effect on the amplitudes requires a better understanding of the freedom to include derivatives in field redefinitions, which we leave for future work.

We see that, at tree level, the energy bound inferred from the four-point amplitude scales as $\sim 4 \pi \mathcal{K}_h^{-\sfrac12}$, which can differ from that of the $n$-point which scales as $\sim 4 \pi v_\star$; in examples of models in the ``weakly curved'' regime given in \cref{sec:UVCompletions}, the bound inferred using the four-point amplitude alone is parametrically weaker. However, certain four-point amplitudes will always ``know'' about the $4 \pi v_\star$ cutoff at loop level. Through the optical theorem \cref{eq:OpticalTheorem}, if \cref{eq:FRredux} violates perturbative bounds, then the imaginary part of $\amp(\pi\pi\to\pi\pi)$ must as well.

\section{HEFT Violates Unitarity at $4\pi v$}
\label{sec:UVCompletions}

In the previous section, we derived a unitarity cutoff of $4 \pi v_\star$ for any HEFT Lagrangian, where the radius of convergence $v_\star$ is the distance (in the complex plane of $\canonh$) between the vacuum and a non-analyticity in a curvature invariant. In this section, we study a series of example UV completions to elucidate the physical properties of such a pole by revealing how it arises from integrating out UV states that unitarize the amplitudes.

In previous work \cite{Cohen:2020xca}, we identified two classes of UV completion where HEFT was required to describe the IR.  One, UV completions containing states that get most of their mass from electroweak symmetry breaking; we will call this type of particle a ``Loryon,'' following \cite{Banta:2021dek}. Two, those containing electroweak charged UV states that provide extra sources of electroweak symmetry breaking, such that the vacuum configuration breaks electroweak symmetry, even when the Higgs vev is turned off. In both cases, the low-energy dynamics of the scalar sector is not well-described by SMEFT, in the sense that the resulting effective action does not admit a convergent expansion -- a local EFT -- in terms of fields linearly realizing electroweak symmetry, such as the Higgs doublet $H$.

Below, we consider representative examples of each class of UV completion and show how, in theories that are poorly described by SMEFT, the scale $v_\star$ must be order the electroweak scale $v \sim G_\text{F}^{-\sfrac12}$. The unitarity cutoff of such theories, as well as the masses of the UV states that unitarize them, cannot be made arbitrarily high; they must appear at the TeV scale or below. This connects the geometric classification of \cite{Cohen:2020xca} back to recent results on the TeV-scale cutoffs of non-SMEFT-like theories~\cite{Chang:2019vez,Falkowski:2019tft}.

\subsection{Tree-level Loryon\label{sec:treelevel}}

Our first example generates non-trivial matching coefficients at tree-level, by integrating out a singlet scalar Loryon $S$. This is the same model discussed in \cite[\S6.1]{Cohen:2020xca}. The UV Lagrangian is determined by writing down all renormalizable interactions between $S$ and $H$, assuming that $S$ transforms under a $\mathbb{Z}_2$ symmetry:
\begin{subequations}\label{eq:UVTreeLevelLoryon}
\begin{align}
\mathcal{L}_\text{UV} &= {\left| {\partial H} \right|^2} + \frac{1}{2}\s (\partial S)^2 - V_\text{UV} \,, \\[5pt]
V_\text{UV} &= - \mu_H^2 \s |H|^2 + \lambda_H\s |H|^4 + \frac{1}{2} \big(\s m^2 + \kappa\s |H|^2\s \big)\s S^2 + \frac{1}{4}\s \lambda_S\s S^4 \,. \label{eq:VUVtreelevelLoryon}
\end{align}
\end{subequations}
For the potential to be bounded from below, we require $\lambda_H,\lambda_S>0$ and $4\s\lambda_H\lambda_S>\kappa^2$. We also need $\mu_H^2>0$ so that the minimum of $V_\text{UV}$ break electroweak symmetry. In addition, we assume that $m^2<0$ and $\kappa < 0$ (such that $S$ has a non-zero vev) in order to obtain a non-trivial tree-level EFT.  To derive the matching, we simply solve for $S_\mathbf{c}$, the solution to the classical equation of motion for the Loryon, plug it into the UV Lagrangian, and expand. In \cite[\S6.1]{Cohen:2020xca}, it was shown that requiring that the solution intersects the global minimum of the UV theory yields
\begin{equation}
S_\mathbf{c} = \bigg( \frac{m^2 + \kappa |H|^2}{-\lambda_S} \bigg)^{1/2} + \mathcal{O}\big(\partial^2\big) \,,
\label{eq:ScTreeLevelLoryon}
\end{equation}
resulting in an EFT of the HEFT form given in~\cref{eq:HEFTLagrangian} with explicit form factors
\begin{subequations}\label{eqn:KFtree}
\begin{align}
K(h) &= \sqrt{1 + \delta \frac{\kappa\s (v+h)^2}{2m^2 + \kappa\s (v+h)^2} } \,,
\qquad\quad  {v}\s F( h ) = v + h \,, \\[10pt]
V(h) &= -\frac12\s \mu_H^2\s (v+h)^2 + \frac14\s \lambda_H (v+h)^4 - \frac{1}{16\s \lambda_S} \Big[ 2\s m^2 + \kappa\s (v+h)^2 \Big]^2 \,,
\end{align}
\end{subequations}
with
\begin{equation}
\delta \equiv -\frac{\kappa}{2\lambda_S} > 0 \qquad \text{and} \qquad v \equiv \sqrt{ \frac{\mu_H^2 - m^2\delta}{\lambda_H + \frac{\kappa}{2}\delta} } \,,
\end{equation}
where the latter ensures that the global minimum of the EFT potential occurs at $h=0$.

Using \cref{eq:SecCurv}, we can calculate the relevant curvature invariants of the EFT manifold
\begin{subequations}\label{eq:treelevelcurvatures}
\begin{align}
\mathcal{K}_h &= \frac{2m^2\kappa\s \delta}{\big[ 2m^2 + \kappa\s (1 + \delta) (v+h)^2 \big]^2} \,, \\[5pt]
\mathcal{K}_\pi &= \frac{\kappa\s \delta}{2m^2 + \kappa\s (1+\delta) (v+h)^2} \,,\\[5pt]
\nabla^2 V - \partial_\canonh^2 V &= 3 \left( \lambda_H  + \frac12 \kappa \delta \right) h (2v + h) \frac{2m^2 + \kappa\s (v+h)^2}{ 2m^2 + \kappa\s (1 + \delta) (v+h)^2 } \,.
\end{align}
\end{subequations}
The three invariants have poles in the complex $h$ plane at\footnote{In the limit $m^2=0$ (the BSM singlet is exactly massless before electroweak symmetry breaking), only $\mathcal{K}_\pi$ retains a non-analyticity; the residues of the poles in the other two invariants would vanish. Thus, only amplitudes with at least four external Goldstones would grow with energy when $m^2 = 0$, see~\cref{eq:FourGoldstoneAmplitudeExplicit}.}
\begin{equation}
h_\star = -v \pm i \sqrt{ \frac{2 m^2}{\kappa\left(1+\delta\right)} } \,.
\label{eq:hstar}
\end{equation}
In order to use this result to determine the unitarity cutoff, we need to transform to the canonically normalized Higgs coordinate $\canonh = Q(h)$ defined in \cref{eq:RealCanonhDefinition}. Note that $K(h)$ remains analytic inside the disk of the radius $\left|h_\star\right|$ centered at the origin $h=0$. So $Q^{-1}(\canonh)$ remains analytic up to the points $\canonh_\star=Q(h_\star)$, which then are the non-analyticities in $\mathcal{K}_{h/\pi}(h)=\mathcal{K}_{h/\pi} (Q^{-1}(\canonh))$ that are closest to the origin $\canonh=0$. Therefore, the radius of convergence of the sectional curvatures in the complex plane of $\canonh$ is
\begin{equation}
v_\star = \big|Q\big(h_\star\big)\big| \,,
\end{equation}
using either root from \cref{eq:hstar}.

\begin{table}[t!]
\renewcommand{\arraystretch}{1.5}
\setlength{\tabcolsep}{0.8em}
\setlength{\arrayrulewidth}{1.2pt}
\centering
\begin{tabular}{c | c c c | c c c}
& $-m^2/v^2$ & $-\kappa/2$ &  $\lambda_S$ & $v_\star/v$ & $v_\star^2\, \overline{\mathcal{K}}_h$ & $v_\star^2\, \overline{\mathcal{K}}_\pi$\\ \hline
  A & $4 \pi$ & $4\pi$ & $8 \pi$ & 1 & 0.1 & 0.3\\
  B &  $0.1$ & $4 \pi$ & $8 \pi$ & 1 & 0.003 & 0.5\\
  C & $0.1$ & $0.1$ & $2$ & 2 & 0.02 & 0.05\\
  D & $10^6$ & $0.1$ & $2$ & 3000 & 0.04 & 0.04
\end{tabular}
\caption{Four benchmark parameter points for exploring the tree-level Loryon model.}
\label{tab:TreeLoryon}
\end{table}

\begin{figure}[t!]
\centering
\hspace{-35pt} \includegraphics[width=0.45\textwidth]{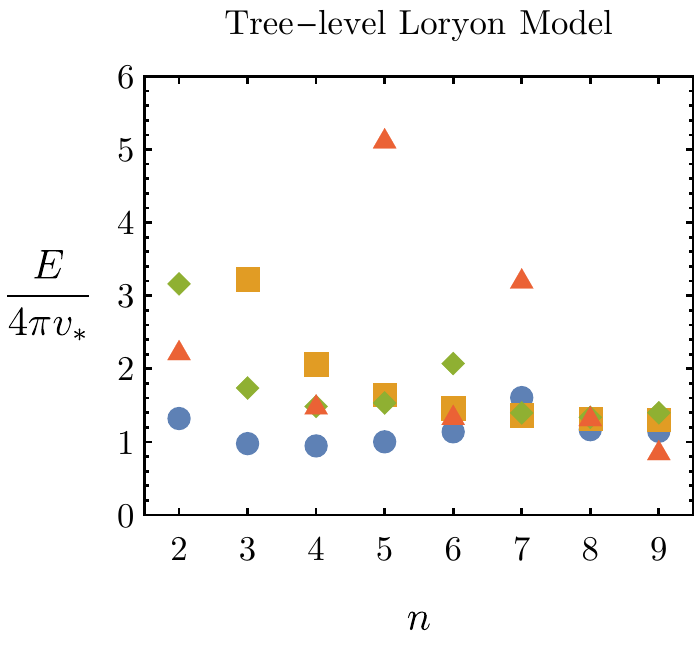}\\
   \includegraphics[align=c,width=0.26\textwidth]{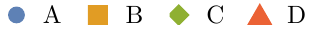}
  \caption{The unitarity cutoff $E$ normalized by $4\pi v_\star$ derived from the process of two Goldstones scattering into $n$ Higgs bosons as computed using \cref{eq:TwoGoldstoneUnitarityBound}, for the four benchmark parameters given in \cref{tab:TreeLoryon}. }
  \label{fig:TreeLevelSingletUnitarityCutoff}
\end{figure}

Having identified the radius of convergence $v_\star$, we can then compute the unitarity cutoff $E$ with \cref{eq:TwoGoldstoneUnitarityBound}. Recall from \cref{eq:EUnitarityBound} that we expect $E$ to converge to $\sim 4 \pi v_\star$ for $n$ of order a few. If true, it confirms our intuition that the scale of unitarity violation can be taken arbitrarily large in the ``SMEFT limit.''  When taking $m^2/v^2 \to \infty$ with $\kappa$ and $\delta$ fixed, \cref{eq:hstar} implies that distance to the pole $h_\star$ grows as $m^2/v^2$. In this region of parameter space, the singlet is getting the minority of its mass from electroweak symmetry breaking. The unitarity cutoff $E \sim 4 \pi v_\star \gg 4 \pi v$, and the SMEFT expansion of the EFT Lagrangian converges rapidly at our vacuum.

To explicitly verify \cref{eq:EUnitarityBound}, and also to explore the regions of parameter space were HEFT is required, we turn to numerics. The primary result of this section is provided in~\cref{fig:TreeLevelSingletUnitarityCutoff}, where we show the behavior of the unitarity cutoff $E$ as we vary the number of Higgs bosons in the final state $n$ for the four benchmark parameter choices given in~\cref{tab:TreeLoryon}. This is obtained by computing each factor in \cref{eq:TwoGoldstoneUnitarityBound} numerically. In particular, the values of $a_n$ for a few choices of $n$, defined in \cref{eq:LimitingScaleSectionalCurvature}, are plotted in \cref{fig:an} for the four benchmark parameter choices given in~\cref{tab:TreeLoryon}. We see from \cref{fig:TreeLevelSingletUnitarityCutoff} that \cref{eq:EUnitarityBound} indeed holds for all four of these benchmarks. Some additional features of each of these benchmark points are:
\begin{itemize}
\item \textbf{Point A:} From~\cref{tab:TreeLoryon} we see that the sectional curvature $\mathcal{K}_h$ is close to $\mathcal{O}(1)$ in units of $v_\star$.  This implies that this HEFT is in the ``strongly curved'' regime. Therefore, the unitarity bound is essentially saturated for $n=2$.
\item \textbf{Point B:} The mass parameter for the Loryon in this example is taken to be much smaller than the mass it acquires from electroweak symmetry breaking. As $v_\star^2\overline{\mathcal{K}}_h\ll 1$, this provides a ``weakly curved'' example of HEFT in the $h$-$\pi_i$ plane, with $v_\star \simeq v$, even though the parameters of the UV theory are large. In this case, the unitarity bound saturates for $n \gtrsim 5$. Note, however, that in the $\pi_i$-$\pi_j$ plane, the manifold is `strongly curved'' ($v_\star^2\, \overline{\mathcal{K}}_\pi \sim 0.5$) so this behavior is not manifest in the four-Goldstone amplitudes.
\item \textbf{Point C:} This provides an example where the UV theory parameters are in the perturbative regime, and the HEFT description is ``weakly curved.'' As compared to Point B, the value of $v_\star$ is higher by a factor of $\sim 2$, and the unitarity bound is saturated for $n \gtrsim 3$ due to $v_\star^2 \, \overline{\mathcal{K}}_h$ being an order of magnitude larger.
\item \textbf{Point D:} This point illustrates the decoupling limit, \emph{i.e.}, $m^2$ is taken to be large, so that the model can be matched onto SMEFT. The scale of unitarity violation $v_\star$ is becoming large, and the unitarity bound is converging for $n \gtrsim 8$.
\end{itemize}

This completes our numerical study of this model. Before moving on to the loop-level example in \cref{sec:looplevel}, we will briefly discuss the role of the Loryon in restoring the apparent unitarity violation derived by studying HEFT alone.

\subsubsection{Who Restores Unitarity?}

Given some simplifying assumptions about the UV parameters, we can analytically explore the connection between $v_\star$ and the mass of the BSM singlet state we have integrated out. This will make it clear that ``integrating in'' this state yields unitary scattering amplitudes in the UV theory. We take the parameter $\delta\equiv-\kappa/(2\lambda_S)$ to be small, and specifically work in the limit that it goes to zero.\footnote{Note that if we take $\delta = 0$ precisely, this would correspond to $\lambda_S \to \infty$, which invalidates this perturbative analysis. Including finite $\delta$ does not change the qualitative nature of this argument.} In this case, we have
\begin{subequations}
\begin{align}
h_\star &= -v \pm i \sqrt{ \frac{2 m^2}{\kappa} } + \mathcal{O}(\delta) \,, \label{eqn:hstarLO} \\[8pt]
K(h) &= 1 + \mathcal{O}(\delta)   \quad\Longrightarrow\quad   Q(h) = h + \mathcal{O}(\delta) \,, \\[8pt]
v_\star^2 &= \big|Q\big(h_\star\big)\big|^2 = \big|h_\star\big|^2 + \mathcal{O}(\delta) = v^2 + \frac{2 m^2}{\kappa} + \mathcal{O}(\delta) \,,
\label{eq:vStarExpanded}
\end{align}
\end{subequations}
Therefore, the locations of the non-analyticity (given by \cref{eqn:hstarLO}) are in one-to-one correspondence with the points on the manifold where the singlet $S$ in the UV theory becomes massless:\footnote{There is one limiting case that is not covered by this analysis, namely if one takes $\kappa \to 0$ and $m^2 \to 0$ with $m^2/\kappa$ fixed.  The Loryon becomes massless in this limit, but the analysis here makes it seem that the scale of unitarity violation is still $\sim 4\pi v$.  This apparent issue would be resolved by including the contribution from the four derivative terms in the analysis.}
\begin{equation}
\frac{\partial^2 V_\text{UV}}{\partial S^2} \bigg\vert_{S=S_\mathbf{c}} = m^2 + \kappa \left|H\right|^2 + 3\lambda_S S_\mathbf{c}^2 = -2 m^2 - \kappa\, (v+h)^2 \,,
\end{equation}
where $V_\text{UV}$ is given in \cref{eq:VUVtreelevelLoryon}, and $S_\mathbf{c}$ is given in \cref{eq:ScTreeLevelLoryon}. Defining the fraction of the mass squared that the singlet gets from electroweak symmetry breaking\footnote{We emphasize that $f$ is a dimensionless fraction here, not to be confused with the scale of global symmetry breaking that appears in composite Higgs models.}
\begin{equation}
f \equiv \frac{ \frac12 \kappa v^2}{m^2 + \frac12 \kappa v^2} \,,
\label{eq:FracEWSBMass}
\end{equation}
we can rewrite the radius of convergence of the sectional curvatures about the origin given in \cref{eq:vStarExpanded}:
\begin{equation}
v_\star^2  = \frac{v^2}{f} \,.
\end{equation}
If the singlet gets the majority of its mass from electroweak symmetry breaking, then $f > 1/2$ and $v_\star \sim v$.

If the UV theory is given by~\cref{eq:UVTreeLevelLoryon}, then the amplitudes $\amp\left(\pi_i \pi_j \to h^n\right)$ considered here are ultimately unitarized by the inclusion of these singlet states as propagating degrees of freedom.  The threshold for producing a on-shell singlet states is $E \geq \left( -2 m^2 - \kappa v^2 \right)^{1/2} = (-\kappa)^{1/2} v_\star$.  Noting that there is also a unitarity bound on the coupling $\kappa \lesssim 4 \pi$, as can be computed within the UV theory, we see that this threshold is at or below the unitarity cutoff scale $4 \pi v_\star$.

\begin{figure}[t!]
\begin{center}
  \begin{tikzpicture}[scale=2.2]
    \draw[->] (-2,0) -- (0.5,0) node[right] {$\Re h$};
    \draw[->] (0,-0.5) -- (0,2) node[right] {$\Im h$};
    \draw[blue,very thick] ({1.5*cos(80)},{1.5*sin(80)}) arc (80:190:1.5);
    \draw[orange,very thick] ({0.75*cos(-10)+1.5*cos(150)},{0.75*sin(-10)}) arc (-10:190:0.75);
    \draw ({1.5*cos(150)},0) -- ({1.5*cos(150)},-0.2) node[below] {$-v$};
    \draw (0,{1.5*sin(150)}) -- (0.2,{1.5*sin(150)}) node[right] {$\displaystyle \sqrt{\frac{2 m^2}{\kappa}}$};
    \node[label={[label distance=-12pt]135:{$h_\star$}}] at ({1.5*cos(150)},{1.5*sin(150)}) {\Large$\star$};
    \draw[->] (0,0) -- ({1.5*cos(150)},{1.5*sin(150)}) node[midway,above,sloped] {$\displaystyle  |h_\star| = \frac{v}{\sqrt{f}}$};
    \node[orange] at (-2.2,0.8) {SMEFT};
    \node[blue] at (-1.0,1.5) {HEFT};
  \end{tikzpicture}
  \end{center}
\caption{An illustration of the complex $h$ plane for the tree-level Loryon model in the limit $\delta \to 0$. The location of the non-analyticity in the sectional curvatures $\mathcal{K}_h$, $\mathcal{K}_\pi$, and the potential terms is denoted by $h_\star$, see~\cref{eqn:hstarLO}.
The location of this non-analyticity sets the radius of convergence for the Taylor expansions of the sectional curvatures about the physical vacuum $h=0$ (\textcolor{blue}{HEFT}, shown in \textcolor{blue}{blue}) and about the electroweak preserving vacuum $h=-v$ (\textcolor{orange}{SMEFT}, shown in \textcolor{orange}{orange}). When the singlet gets the majority of its mass from electroweak symmetry breaking, then $f > 1/2$ and the SMEFT expansion does not converge at the origin.  This implies that SMEFT cannot describe the UV model observables arbitrarily well.}
  \label{fig:RadiusOfConvergence}
\end{figure}
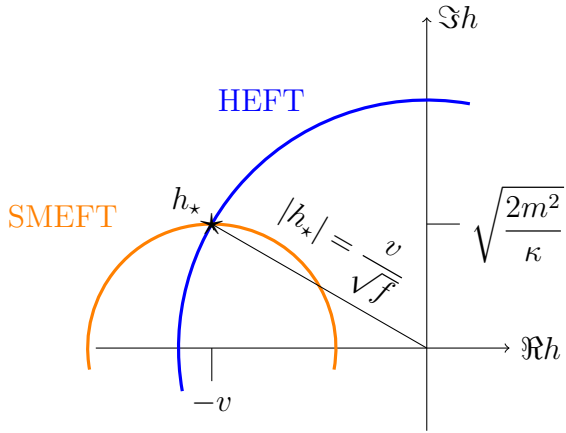

This leads us to the intuitive situation illustrated in \cref{fig:RadiusOfConvergence}, where we have sketched the complex $h$ plane. The electroweak preserving vacuum is located at the point $\Re h = -v$, and $h_\star$ is the location of the non-analytic point (a pole in this example).  This pole sets the radius of convergence for the two EFT expansions, where SMEFT (HEFT) is centered about the point where electroweak symmetry is restored (the physical vacuum).  If SMEFT is to provide a convergent expansion about the physical vacuum, then the origin $h=0$ must lie within the expansion's circle of convergence; this only occurs if $f < 1/2$.  For contrast, the situation shown in \cref{fig:RadiusOfConvergence} is the case when $f > 1/2$, the SMEFT expansion does not converge at the physical vacuum (the orange circle does not enclose the origin).  This implies that one must use HEFT to describe this UV model at low energies, with a corresponding unitarity cutoff set by $4 \pi v_\star \sim 4 \pi |h_\star| \sim 4 \pi v$.

\subsection{One-loop Loryon\label{sec:looplevel}}

Next, we consider the IR limit of the $\mathbb{Z}_2$ symmetric phase of the singlet model defined in~\cref{eq:UVTreeLevelLoryon}.  Specifically, we are working in the parameter space where $m^2 > 0$ and $\kappa > 0$, such that the tree-level solution $S_\mathbf{c}=0$ is at the global minimum of the theory. Tree-level matching does not yield any deviations from the Standard Model, and so we focus on matching this UV theory onto an EFT at one-loop order. This is the same model discussed in~\cite[\S6.2]{Cohen:2020xca}.

We use functional methods to integrate out fluctuations around the trivial tree-level solution to obtain the one-loop effective action, which results in the form factors
\begin{subequations}
\begin{align}
K(h) &= \sqrt{ 1 + \frac{\kappa}{96\s\pi^2} \frac{\kappa\s(v+h)^2}{2\s m^2 + \kappa\s(v+h)^2}} \,, \\[8pt]
vF(h) &= v + h \,.
\end{align}
\end{subequations}
In addition to these two form factors, one must also compute the one-loop contributions to $V_{\text{eff}}$, which yields the familiar Coleman-Weinberg potential for the light fields; the explicit expression is given in~\cite[\S6.2]{Cohen:2020xca}. We note that the kinetic term form factors above have the same functional form as in \cref{eqn:KFtree}. We find the same sectional curvatures for this EFT manifold as in~\cref{eq:treelevelcurvatures} if we make the substitution $\delta \mapsto \kappa/(96 \pi^2)$, together with a potential term:
\begin{subequations}\label{eq:looplevelcurvatures}
\begin{align}
\mathcal{K}_h &= \frac{m^2\kappa^2}{48\pi^2\big[ 2m^2 + \kappa\s \big(1 + \frac{\kappa}{96\pi^2}\big) (v+h)^2 \big]^2} \,, \\[10pt]
\mathcal{K}_\pi &= \frac{\kappa^2}{96\pi^2\big[2m^2 + \kappa\s \big(1 + \frac{\kappa}{96\pi^2}\big) (v+h)^2\big]} \,, \\[10pt]
\nabla^2 V - \partial_\canonh^2 V &= 3\, \frac{2m^2 + \kappa\s (v+h)^2}{ 2m^2 + \kappa\s (1 + \frac{\kappa}{96\pi^2}) (v+h)^2 } \Bigg\{ -\mu_H^2 + \lambda_H (v+h)^2 \notag\\[5pt]
                                  & \hspace{30pt} - \frac{\kappa}{64 \pi^2} \left[ 2m^2 + \kappa\s (v+h)^2 \right] \left( \ln \frac{\mu^2}{m^2 + \frac12 \kappa\s (v+h)^2}+ 1 \right) \Bigg\} \,.
\end{align}
\end{subequations}
The above curvature invariants have poles in the complex-$h$ plane at
\begin{equation}
h_\star = -v \pm i \sqrt{ \frac{2 m^2}{\kappa\left(1+ \frac{\kappa}{96\pi^2}\right)} } \,.
\label{eq:hstarOneLoop}
\end{equation}
Since only the $\kappa$ coupling is relevant to the one-loop matching, the shape of the EFT manifold is fully determined at this order by the two UV parameters, $m^2$ and $\kappa$.\footnote{In the symmetric phase, the UV parameter $\lambda_S$ only enters at two-loop order.}  Taking $m^2/v^2 \gg 1$ yields the ``SMEFT limit'' in exactly the same way as described in \cref{sec:treelevel} above.

\begin{figure}[t!]
\centering
\hspace{-35pt}\includegraphics[width=0.45\textwidth]{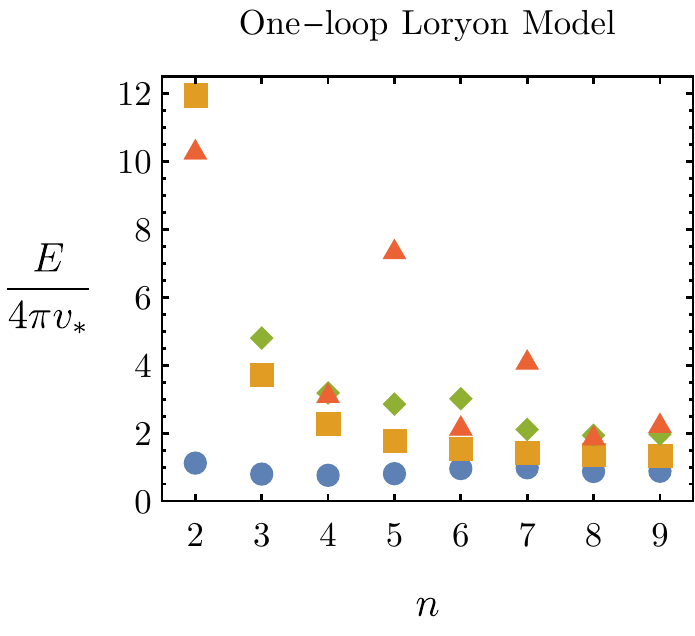}\\
   \includegraphics[align=c,width=0.26\textwidth]{figs/SingletLegend.pdf}
  \caption{The unitarity cutoff $E$ normalized by $4\pi v_\star$ derived from the process of two Goldstones scattering into $n$ Higgs bosons as computed using \cref{eq:TwoGoldstoneUnitarityBound}, for the four benchmark parameters given in \cref{tab:LoopLoryon}. }
  \label{fig:LoopLevelSingletUnitarityCutoff}
\end{figure}

For completeness, we have provided four numerical benchmark points for this model in \cref{tab:LoopLoryon}. The associated $a_n$ plots are given in \cref{fig:an}, and the unitarity cutoff as a function of the number of final state Higgs bosons $n$ is given in \cref{fig:LoopLevelSingletUnitarityCutoff}.  The reasons for choosing each benchmark are identical to  \cref{sec:treelevel}, and so we will succinctly summarize their behavior here.
\begin{itemize}
\item \textbf{Point A:} This shows HEFT is in the ``strongly curved'' regime, and the unitarity bound is essentially already saturated for $n=2$.
\item \textbf{Point B:}  This is a ``weakly curved'' example of HEFT with $v_\star^2\overline{\mathcal{K}}_h \ll 1$ and $v_\star \simeq v$; the unitarity bound saturates for $n \gtrsim 5$. Note that its other sectional curvature $v_\star^2\, \overline{\mathcal{K}}_\pi \sim 0.3$ is much larger, meaning that it is ``strongly curved'' in the $\pi_i$-$\pi_j$ plane, which will be reflected by the behavior of four-Goldstone amplitudes.
\item \textbf{Point C:}  This is another ``weakly curved'' HEFT example, where now both the mass parameter and the coupling to the Higgs are taken to be smaller.
\item \textbf{Point D:}  This point illustrates the decoupling limit, \emph{i.e.}, $m^2$ is taken to be large, so that the model can be matched onto SMEFT.  The scale of unitarity violation $v_\star$ is becoming large, as can be seen from~\cref{tab:LoopLoryon}.
\end{itemize}

\begin{table}[t!]
\renewcommand{\arraystretch}{1.5}
\setlength{\tabcolsep}{0.8em}
\setlength{\arrayrulewidth}{1.2pt}
\centering
  \begin{tabular}{c | c c | c c c }
     &  $m^2/v^2$ & $\kappa/2$ & $v_\star/v$ & $v_\star^2 \,\overline{\mathcal{K}}_h$ & $v_\star^2 \,\overline{\mathcal{K}}_\pi$\\ \hline
    A &  $(4\pi)^2$ & $(4 \pi)^2$ & 2 & 0.2 & 0.4\\
    B &  $1$ & $(4 \pi)^2$ & 1 & $1\times 10^{-3}$ & 0.3\\
    C & $1$ & $1$ & 1 & $1\times 10^{-3}$ & $2\times 10^{-3}$ \\
    D & $10^6$ & $1$ &  1000 & $2\times 10^{-3}$ & $2\times 10^{-3}$
  \end{tabular}
  \caption{Four benchmark parameter points for exploring the one-loop Loryon model.}
  \label{tab:LoopLoryon}
  \end{table}

This completes our discussion of the behavior of a model whose leading matching coefficients are determined at one-loop order.  Unsurprisingly, we find essentially the same physics as in the tree-level Loryon model discussed in the previous section.  In the next section, we present our final case study: a model with an additional source of spontaneous electroweak symmetry breaking.

\subsection{BSM Symmetry Breaking}

Up to this point, we have explored two phases of a model that involves a BSM state which does not participate in electroweak symmetry breaking.  We saw that in the regions of parameter space where this Loryon received more than half of its mass from the Higgs vev such that the model must be matched onto HEFT, the scale of unitarity violation was $\sim 4\pi v$.  Our goal here is to explore the other class of model that requires HEFT discussed in~\cite{Cohen:2020xca}, one that involves additional sources of electroweak symmetry breaking.

Due to the inherent complexities for models involving fields with symmetry breaking vevs, we will choose to simplify this analysis by studying the Two Abelian Higgs model discussed in~\cite[\S7.1]{Cohen:2020xca}.  This serves as a toy model for the situation in which fields that are associated with extra sources of symmetry breaking are integrated out.  We introduce two complex scalars, $H_A$ and $H_B$, which have respective charges $+2$ and $+1$ under a global $SO(2)$ symmetry, which acts as a proxy for the custodial $O(4)$ symmetry of the SM scalar sector. The UV Lagrangian is given by
\begin{subequations}
\begin{align}
\mathcal{L}_\text{UV} &= {\left| {\partial {H_A}} \right|^2} + {\left| {\partial {H_B}} \right|^2} - V_\text{UV} \,, \\[10pt]
V_\text{UV} &= m_A^2\s | H_A |^2 + m_B^2\s | H_B |^2 + \lambda_A\s | H_A |^4 + \lambda_B\s | H_B |^4 + 2\s \kappa\s  | H_A |^2  | H_B |^2 \notag\\[5pt]
&\quad + \Big[ \mu\s H_A\s \big(H_B^*\big)^2 + \text{h.c.} \Big] \,.
\end{align}
\label{eq:AbelianHiggsUVLinearBasis}
\end{subequations}
For the potential to be bounded from below, we require $\lambda_A,\lambda_B>0$ and $\lambda_A\s \lambda_B > \kappa^2$.

Our goal is to integrate out $H_A$.  To this end, it is convenient to parametrize the fields in the ``unitary basis''~\cite{Cohen:2020xca}:
\begin{subequations}\label{eqn:repara}
\begin{align}
H_B &= \frac{1}{\sqrt{2}}\, r\s e^{i\s \pi} = \frac{1}{\sqrt{2}}\, (v+h)\s e^{i\s \pi} \,, \\[5pt]
H_A &= \sqrt{2}\,\frac{f}{r^2}\, e^{i\s \beta}\s H_B^2 = \frac{1}{\sqrt{2}}\, f\s e^{i\s \beta + 2\s i\s \pi } \,.
\end{align}
\end{subequations}
Note that we have made the identification $r \equiv v + h$ such that the light scalar field modulus square is $\left|H_B\right|^2 = \frac12 r^2 = \frac12 (v+h)^2$, analogous to our four-component Higgs studied in the previous two examples.

In order to obtain the tree-level EFT Lagrangian, we need to solve for the solutions to the equations of motion for $\beta$ and $f$:
\begin{subequations}
\begin{align}
\beta_\mathbf{c} &= -\arg \mu + \pi \,, \\[5pt]
f_\mathbf{c} &= \left( q + \Delta^{1/2} \right)^{1/3} + \left( q - \Delta^{1/2} \right)^{1/3} \,,
\label{eq:fsol}
\end{align}
\end{subequations}
where $\pi$ is the Archimedes' constant, $\pi=3.14159\ldots$, not to be confused with the field $\pi$ introduced in \cref{eqn:repara}, and
\begin{subequations}
\begin{align}
\Delta(r) &= q^2 + p^3 \,, \\[8pt]
p(r) &= \frac{m_A^2+\kappa\s r^2}{3 \lambda_A} \,, \\[5pt]
q(r) &= \frac{ |\mu| r^2}{\sqrt{8} \lambda_A} \,.
\end{align}
\end{subequations}
Note that when interpreting \cref{eq:fsol}, we take the cube roots to be on the principal branch, such that $f_\mathbf{c}$ is real (and positive) for all real $r$. This guarantees that the resulting EFT includes the global minimum of the UV theory.

In terms of $f_\mathbf{c}$, the EFT Lagrangian is\footnote{Note that there is no $\mathcal{K}_\pi$ sectional curvature in this model, since the symmetry group is $SO(2)$.}
\begin{equation}
\mathcal{L}_\text{EFT} = \frac12 \Big[ 1 +  ( f_\textbf{c}^\prime)^2 \Big] (\partial r)^2 + \frac12 \left( r^2 + 4 f_\textbf{c}^2 \right) (\partial \pi)^2 - V_\text{EFT} \,,
\end{equation}
where a prime denotes the differentiation with respect to $r$. The resulting sectional curvature of the 2D EFT manifold (charted by $r$ and $\pi$) is
\begin{align}
\mathcal{K}_h &= -\frac{4 \left(f_\mathbf{c} - r f_\mathbf{c}^\prime \right)^2 \left( 1 + (f_\mathbf{c}^\prime)^2 \right) + f_\mathbf{c}^{\prime\prime} \left( 4 f_\mathbf{c} - r f_\mathbf{c}^\prime \right) \left( r^2 + 4 f_\mathbf{c}^2 \right)}{\big( r^2 + 4 f_\mathbf{c}^2 \big)^2 \big( 1 + (f_\mathbf{c}^\prime)^2 \big)^2} \,.
\label{eq:AbelHiggsSectionalCurvature}
\end{align}
From this formula, we see that the function $\mathcal{K}_h$ defined in the complex $r=v+h$ plane can only have non-analyticities when
\begin{enumerate}
  \item $1 + (f_\mathbf{c}^\prime)^2 = 0$,
  \item $r^2 + 4 f_\mathbf{c}^2 = 0$,
  \item $f_\mathbf{c}$ is itself non-analytic.
\end{enumerate}

\begin{figure}[t!]
\centering
\hspace{-35pt}\includegraphics[width=0.45\textwidth]{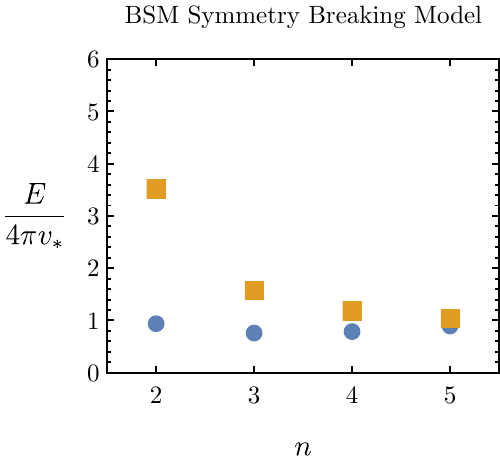}\\
   \includegraphics[align=c,width=0.13\textwidth]{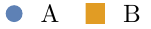}
  \caption{The unitarity cutoff $E$ normalized by $4\pi v_\star$ derived from the process of two Goldstones scattering into $n$ Higgs bosons as computed using \cref{eq:TwoGoldstoneUnitarityBound}, for the two benchmark parameters given in \cref{tab:AbelianHiggsTable}.}
  \label{fig:AbelHiggsUnitarityCutoff}
 \end{figure}

A thorough exploration of how each of these conditions manifests on the parameter space is beyond the scope of this work.  In order to simply show that the lessons we learned from the above studies hold in the case when there is BSM symmetry breaking, we appeal to numerics.  In particular, we have provided two benchmark parameter points in~\cref{tab:AbelianHiggsTable}, and the associated $a_n$  values are given in~\cref{fig:an}.   For both benchmarks, the non-analyticity in $\mathcal{K}_h$ that is closest to the vacuum occurs when $1 + (f_\mathbf{c}^\prime)^2 = 0$.  We use this to compute the radius of convergence $v_\star = \left|Q(h_\star)\right|$ numerically from its location. The resulting unitarity cutoff is depicted in \cref{fig:AbelHiggsUnitarityCutoff} for these two benchmark points, both with $v_\star \simeq v$.

\begin{itemize}
\item \textbf{Point A:} This shows HEFT is in the ``strongly curved'' regime, and the unitarity bound is essentially already saturated for $n=2$.
\item \textbf{Point B:}  This is a ``weakly curved'' example of HEFT with $v_\star \simeq v$, and the unitarity bound saturates for $n \gtrsim 5$.
\end{itemize}

\begin{table}[t!]
\renewcommand{\arraystretch}{1.5}
\setlength{\tabcolsep}{0.8em}
\setlength{\arrayrulewidth}{1.2pt}
\centering
  \begin{tabular}{c | c c c c c c | c c}
       & $m_A^2/v^2$ & $m_B^2/v^2$ & $\mu/v$ & $\lambda_A$ & $\lambda_B$   & $\kappa$ & $v_\star/v$ & $v_\star^2\, \overline{\mathcal{K}}_h$ \\ \hline
    A & -0.43 &-0.43  & 0.29 & 0.7 & 0.7 & 0.1 & 1.3 & 0.26 \\
    B &  -0.069 & -0.69 & 0.035 & 7 & 0.7 & 0.01 & 1.0 & 0.018
  \end{tabular}
\caption{Two benchmark parameter points for exploring the Two Abelian Higgs model.}
\label{tab:AbelianHiggsTable}
\end{table}

This example demonstrates that when one is matching a UV model that includes additional sources of spontaneous symmetry breaking onto HEFT, the unitarity bound is $\lesssim 4 \pi v$ as anticipated.  Although this analysis was performed on a toy model, the same conclusions are expected to hold for more realistic examples such as the Two Higgs Doublet Model or the Triplet Higgs extension of the Standard Model, which were both discussed in~\cite{Cohen:2020xca}.

\section{Conclusions}
\label{Sec:Conc}

How can we probe the geometry of an Effective Field Theory? In this work, we have explored the sense in which scattering amplitudes measure the geometry of the scalar manifold in EFTs of the Higgs sector. We began by obtaining general expressions for $n$-point amplitudes involving the Higgs scalar $h$ and Goldstone bosons $\pi_i$ in terms of the sectional curvatures, scalar potential, and covariant derivatives thereof. Focusing on the high-energy behavior of these amplitudes, we linked the geometric classification of HEFT and SMEFT to more familiar unitarity-based arguments, connecting the presence of non-analyticities in curvature invariants to the scale of unitarity violation. In particular, we demonstrated that when these non-analyticities are sufficiently close to a putative fixed point -- such that HEFT is required by geometric criteria -- the scale of unitarity violation in sufficiently high-multiplicity amplitudes is $\lesssim 4 \pi v$.

This provides the missing link among classifications of HEFT based respectively on unitarity, analyticity, and geometry. The key observation is that, whereas 2-to-2 amplitudes only measure the curvature of the scalar manifold at our vacuum, higher-point amplitudes begin to reconstruct the curvature further afield. This resonates with recent results on probing the scale of unitarity violation with higher-point amplitudes in HEFT \cite{Chang:2019vez,Falkowski:2019tft, Abu-Ajamieh:2020yqi}. In addition to presenting general arguments linking geometry, analyticity, and unitarity, we applied these arguments in the context of several concrete BSM examples where the non-analyticities are directly associated with additional degrees of freedom. Beyond the applications to unitarity and geometry pursued here, our basis-independent expressions for scalar scattering amplitudes in terms of geometric invariants may prove more broadly useful to studies of SMEFT and HEFT, particularly as a bridge between Lagrangian parameterizations and purely on-shell formulations \cite{Shadmi:2018xan, Durieux:2019eor, Durieux:2020gip}. It is also necessary to include fermions and vectors, for which there exist prescriptions for encoding their EFT Lagrangians as covariant objects on target manifolds (see, \eg, \cite{Finn:2020nvn,Nagai:2019tgi}). It will be fascinating to study how the normal coordinate expansions of such objects map onto the contact terms in amplitudes, analogous to the scalar case presented here.

We emphasize that the amplitudes studied in our work only probe the geometry of the scalar manifold in a local sense, by reconstructing the Taylor expansion of sectional curvatures about our vacuum. While this provides enough information about far-away points on the manifold (such as the possible fixed point) to connect geometry and unitarity, it cannot capture global properties. Doing so presumably involves non-perturbative phenomena, and remains an interesting open question for future study.
Similarly, we emphasize that the presented connection to non-decoupling UV physics has relied on the study of UV models that can be perturbatively matched onto the EFT, yielding functional forms in the EFT Lagrangian with nearby singularities. However, the non-decoupling behavior of HEFT extends beyond perturbation theory: Ref.~\cite{Alonso:2021rac} presents semi-classical arguments that the unitarity cutoff of manifolds lacking an electroweak-symmetry-preserving fixed point cannot be arbitrarily high (irregardless of any singularities in the sectional curvatures).

In a companion work~\cite{Banta:2021dek}, we have initiated a complementary program to systematically characterize perturbative BSM models with new particles (the ``Loryons'') that must be matched onto HEFT. Since at least some of the BSM states in these models must have mass $\lesssim 4 \pi v \simeq 3 \text{ TeV}$, it makes sense to ask what constraints (both direct and indirect) exist for these models.  As we will show in \cite{Banta:2021dek}, there are models with open parameter space that provide concrete targets for searches for both direct production of the Loryons and indirect effects on the electroweak sector as modeled using HEFT.  This presents an exciting opportunity to explore BSM physics that could have a dramatic impact on our understanding of phenomena at the electroweak scale.

Ultimately, this and other work exploring EFT extensions of the Standard Model are all in service of finding new observable phenomena that can be searched for at the LHC or other experiments, while also providing a framework for interpreting null results.  The robust result of this work is that theories whose low-energy physics is described by HEFT must violate unitarity at a scale $\lesssim 4 \pi v \simeq 3 \text{ TeV}$, but not necessarily in 2-to-2 processes at leading momentum order. By combining direct searches and electroweak precision data with the high-energy behavior of high-multiplicity final states, the LHC and proposed future colliders are in a position to discover or exclude HEFT. In the event of null results, the precise combination of measurements required to exclude HEFT remains an open question, but one for which a geometric approach is likely to prove fruitful. Even the exclusion of HEFT in the absence of deviations from the Standard Model would be a remarkable result, as it would demonstrate that the known particles alone linearly realize electroweak symmetry -- a property that is often assumed but remains experimentally unverified. This provides added motivation for the high-luminosity LHC program, as well as proposed future colliders probing the Standard Model at the weak scale and beyond.

\acknowledgments

We thank Ian Banta, Spencer Chang, Lance Dixon, Seth Koren, Markus Luty, and Matthew McCullough for useful conversations. The work of N.~Craig is supported by the U.S.~Department of Energy under the grant DE-SC0011702. The work of T.~Cohen and X.~Lu are supported by the U.S.~Department of Energy under grant number DE-SC0011640. D.~Sutherland has received funding from the European Union's Horizon 2020 research and innovation programme under the Marie Skłodowska-Curie grant agreement No.~754496.

\appendix
\section*{Appendices}
\addcontentsline{toc}{section}{\protect\numberline{}Appendices}%
\addtocontents{toc}{\protect\setcounter{tocdepth}{1}}
\section{Normal Coordinates}
\label{app:NormalCoordinates}

Once we have expressed our amplitudes in terms of curvature invariants, they are manifestly invariant under field redefinitions specified in~\cref{eq:CoordRedef}. Therefore, a useful strategy is to work in an advantageous coordinate system at intermediate steps in order to simplify the derivation. To this end, we will work with (Riemann) normal coordinates, which have the defining property that they specify an ``inertial frame'' locally. Normal coordinates transform contravariantly under the class of field redefinitions specified in~\cref{eq:CoordRedef}. This implies that when the Lagrangian is expressed in this basis, the individual Wilson coefficients (partial derivatives of the metric and potential) are composed of covariant derivatives of the potential and Riemann curvature tensor~\cite{Alvarez-Gaume:1981exa}. In this appendix, we provide a brief summary of these facts.

To introduce the normal coordinates, let us begin by working with a generic set of coordinates $\vec\phi$. Without loss of generality, we set the origin $\vec\phi=\vec 0$ at the physical vacuum. Assume the field manifold is smooth in the neighborhood of the origin, so an arbitrary nearby point $\vec\phi$ can be reached by following a unique geodesic starting from the origin,\footnote{Note that this need only be true in the neighborhood of the origin. Generally, the whole manifold may not be geodesically complete.}
which we parameterize by $\lambda$ and denote by $\vec\phi_\text{geo}(\lambda)$:
\begin{equation}
\frac{\dd^2 \phi_\text{geo}^\alpha}{\dd \lambda^2} + \Gamma^\alpha_{\beta\gamma}\Big(\vec\phi_\text{geo}(\lambda)\Big)\, \frac{\dd \phi_\text{geo}^\beta}{\dd \lambda} \frac{\dd \phi_\text{geo}^\gamma}{\dd \lambda} = 0 \,,
\label{eqn:GeodesicEquation}
\end{equation}
with
\begin{align}
\vec\phi_\text{geo}(0) = \vec 0 \,,\qquad \text{and} \qquad
\vec\phi_\text{geo}(1) = \vec\phi \,.
\end{align}
Each selected point $\vec\phi$ will give us a specific solution $\vec\phi_\text{geo}(\lambda)$, and we can use the unique vector $\vec\eta$ tangent to this geodesic solution at the origin to represent $\vec\phi$:
\begin{equation}
\vec\phi \quad\longrightarrow\quad \vec\eta \equiv \frac{\dd\vec\phi_\text{geo}}{\dd\lambda}(0) \,.
\end{equation}
This vector $\vec\eta$ is the (Riemann) normal coordinate.

Using the normal coordinate $\vec\eta$, the solution to \cref{eqn:GeodesicEquation} can be constructed order-by-order in $\lambda$:
\begin{equation}
\phi_\text{geo}^\alpha(\lambda) = \lambda \, \eta^\alpha - \sum_{n=2}^\infty \frac{1}{n!}\, \lambda^n \, \overline\Gamma^\alpha_{(\beta_1 \dots \beta_n)}\, \eta^{\beta_1} \ldots \eta^{\beta_n} \,,
\label{eq:GeodesicSolution}
\end{equation}
and hence we obtain the explicit map from the normal coordinates $\vec\eta$ to the generic coordinate $\vec\phi$:
\begin{equation}
\phi^\alpha = \phi_\text{geo}^\alpha(1) = \eta^\alpha - \sum_{n=2}^\infty \frac{1}{n!}\, \overline\Gamma^\alpha_{(\beta_1 \dots \beta_n)}\, \eta^{\beta_1} \cdots \eta^{\beta_n} \,.
\label{eq:GenericToNormalCoords}
\end{equation}
In the above, we used the generalized Christoffel symbols $\Gamma^\alpha_{\beta_1 \dots \beta_n}(\vec\phi)$, which are defined recursively from the standard Christoffel symbol by covariant differentiation of the lower indices only
\begin{equation}
\Gamma^\alpha_{\beta_1 \cdots \beta_n \gamma} \equiv
\Gamma^\alpha_{\beta_1 \cdots \beta_n, \gamma} - \sum_{j=1}^n \Gamma^\rho_{\gamma \beta_j} \Gamma^\alpha_{\beta_1 \dots \hat \beta_j \rho \dots \beta_n} \,.
\label{eqn:GeneralChrisDef}
\end{equation}
We also used the bar notation to denote quantities evaluated at the origin
\begin{equation}
\overline \Gamma^\alpha_{\beta_1 \cdots \beta_n} \equiv \Gamma^\alpha_{\beta_1 \cdots \beta_n}\big(\vec\phi=\vec 0\,\big) \,.
\end{equation}

We pause here to note an important property of normal coordinates: rescaling $\vec\eta$ can be interpreted as rescaling $\lambda$ in \cref{eq:GeodesicSolution}, so it simply moves the point along the same geodesic. Therefore all the points on a same geodesic must lie along a straight line when written in normal coordinates; this is a feature by design. In other words, the geodesic equation in~\cref{eqn:GeodesicEquation} must be trivialized when written in the normal coordinates. One useful implication of this is immediately clear from~\cref{eq:GenericToNormalCoords}: assume that the coordinates $\vec\phi$ were already normal coordinates, then in order for \cref{eq:GenericToNormalCoords} to hold, we have
\begin{equation}
\overline\Gamma^\alpha_{(\beta_1 \cdots \beta_n)} \normalcoordssuffix = 0  \qquad \forall n \geq 2 \,.
\label{eqn:SymGenChristoffelIsZero}
\end{equation}
This is a powerful constraint that has many simplifying consequences, providing various advantages to using normal coordinates. For example, taking the $n=2$ case, we find
\begin{equation}
\overline\Gamma^\alpha_{\beta_1\beta_2} \normalcoordssuffix = \overline\Gamma^\alpha_{(\beta_1\beta_2)} \normalcoordssuffix = 0 \,.
\end{equation}
This also implies
\begin{equation}
2\, \overline\Gamma_{\alpha\beta_1\beta_2} \normalcoordssuffix = 2\, \overline{g}_{\alpha\lambda}\, \overline\Gamma^\lambda_{\beta_1\beta_2} \normalcoordssuffix = \left( \overline{g}_{\alpha\beta_1,\beta_2} + \overline{g}_{\alpha\beta_2,\beta_1} - \overline{g}_{\beta_1\beta_2,\alpha} \right) \normalcoordssuffix = 0 \,.
\label{eqn:g3sum}
\end{equation}
Symmetrizing all the indices in \cref{eqn:g3sum} gives us
\begin{equation}
\overline{g}_{(\alpha\beta_1,\beta_2)} \normalcoordssuffix = 0 \quad\Rightarrow\quad
\left( \overline{g}_{\alpha\beta_1,\beta_2} + \overline{g}_{\alpha\beta_2,\beta_1} + \overline{g}_{\beta_1\beta_2,\alpha} \right) \normalcoordssuffix = 0 \,.
\end{equation}
Combining this with \cref{eqn:g3sum}, we see that any first order partial derivative of the metric evaluated at the origin must vanish:
\begin{equation}
\overline{g}_{\alpha\beta_1,\beta_2} \normalcoordssuffix = 0 \,.
\label{eqn:gFirstDerivativeZero}
\end{equation}

Using \cref{eqn:SymGenChristoffelIsZero} beyond the lowest order, one can derive a series of similar simplification features in normal coordinates. Let us summarize a few of them that are relevant for our discussions in this paper. For the Christoffel symbols, one can show that
\begin{subequations}\label{eqn:ChristoffelSymbol}
\begin{align}
\overline\Gamma^\alpha_{(\beta_1\cdots\beta_n,\beta_{n+1}\cdots\beta_{n+k})} \normalcoordssuffix &= 0
\qquad \forall\; n\ge 2 \;\text{ and }\; k\ge 0 \,, \label{eqn:ChristoffelSymbol1} \\[8pt]
\overline\Gamma_{\alpha(\beta_1\beta_2,\beta_3\cdots\beta_{k+2})} \normalcoordssuffix &= 0
\qquad \forall\; k\ge 0 \,. \label{eqn:ChristoffelSymbol2}
\end{align}
\end{subequations}
Note that the $k=0$ case of \cref{eqn:ChristoffelSymbol1}, \ie, no partial derivatives appended, is just \cref{eqn:SymGenChristoffelIsZero}. Similarly, \cref{eqn:ChristoffelSymbol2} is a partial-derivative appended case of \cref{eqn:g3sum}. Partial-derivative appended metric components also vanish, once the indices are partially symmetrized:
\begin{equation}
\overline{g}_{\alpha(\beta_1,\beta_2\cdots\beta_{k+2})} \normalcoordssuffix
= \overline{g}_{(\beta_1\beta_2,\beta_3\cdots\beta_{k+2})\alpha} \normalcoordssuffix
= 0 \qquad \forall\; k \ge 0 \,.
\label{eqn:gsymmetrized}
\end{equation}
These of course also imply that the component vanishes upon full symmetrization of indices
\begin{equation}
\overline{g}_{(\alpha\beta_1,\beta_2\cdots\beta_{k+2})} \normalcoordssuffix = 0 \,.
\end{equation}
Note that in the $k=0$ case, the second equation in \cref{eqn:gsymmetrized} results in \cref{eqn:gFirstDerivativeZero}. We will derive \cref{eqn:ChristoffelSymbol,eqn:gsymmetrized} in \cref{appsubsec:derivations}.

When a generic Lagrangian in~\cref{eq:GenLagExpanded} is written in normal coordinates, the Wilson coefficients, \ie, the partial derivatives of the metric and the potential, also take a simpler form, in the way that they connect to covariant quantities. Partial derivatives of the potential just become symmetrized covariant derivatives (which we will also prove in \cref{appsubsec:derivations}):
\begin{equation}
\overline{V}_{,\gamma_1\cdots\gamma_n} \normalcoordssuffix = \overline{V}_{;(\gamma_1\cdots\gamma_n)} \,.
\label{eqn:VNormalCoords}
\end{equation}
Partial derivatives of the metric (without symmetrizing indices) are more involved. For the first few orders, one can explicitly check that the following hold
\begin{subequations}\label{eqn:gNormalCoordsFirstFew}
\begin{align}
\overline{g}_{\alpha\beta,\gamma_1} \normalcoordssuffix &= 0 \,, \\[5pt]
\overline{g}_{\alpha\beta,\gamma_1\gamma_2} \normalcoordssuffix &= \frac23\, \overline{R}_{\alpha (\gamma_1\gamma_2)\beta} \,, \\[5pt]
\overline{g}_{\alpha\beta,\gamma_1\gamma_2\gamma_3} \normalcoordssuffix &= \overline{R}_{\alpha (\gamma_1\gamma_2|\beta;|\gamma_3)} \,, \\[5pt]
\overline{g}_{\alpha\beta,\gamma_1\gamma_2\gamma_3\gamma_4} \normalcoordssuffix &= \frac65\, \overline{R}_{\alpha(\gamma_1\gamma_2|\beta;|\gamma_3 \gamma_4)} + \frac{16}{15}\, \overline{R}_{\alpha (\gamma_1\gamma_2|\rho} \overline{R}^\rho_{|\gamma_3\gamma_4)\beta} \,.
\end{align}
\end{subequations}
The general expression for the $n^\text{th}$ partial derivative of the metric in normal coordinates may be constructed recursively \cite{Muller:1997zk, Hatzinikitas:2000xe}, and they turn out to satisfy
\begin{align}
  \overline{g}_{\alpha\beta,\gamma_1 \dots \gamma_n} \normalcoordssuffix =&\, 2\, \frac{n-1}{n+1}\, \overline{R}_{\alpha (\gamma_1 \gamma_2| \beta;| \gamma_3 \dots \gamma_n)} \nonumber \\[4pt]
                                                                          & \,+ \sum_{k=2}^{n-2} {n \choose k} \frac{k-1}{k+1} \frac{n-k-1}{n-k+1} \frac{ (n-k)^2 + k^2 + n(n+2) }{n(n+1)} \nonumber \\[4pt]
                                                                          & \hspace{9ex} \times \overline{R}_{\alpha (\gamma_1 \gamma_2 | \rho ; |\gamma_3 \dots \gamma_k} \overline{R}^\rho_{\gamma_{k+1} \gamma_{k+2} | \beta; |\dots \gamma_n)} \nonumber \\[2pt]
                                                                          &\,+ \mathcal{O} \big( \s\overline{R}{\s}^{3} \big) \,.
\label{eqn:gNormalCoords}
\end{align}

\subsection{Derivations}
\label{appsubsec:derivations}

In this subsection, we derive \cref{eqn:ChristoffelSymbol,eqn:gsymmetrized,eqn:VNormalCoords} from \cref{eqn:SymGenChristoffelIsZero}. We will drop the specification
\begin{equation}
\text{``}\normalcoordssuffix\text{''}
\end{equation}
to improve the readability of the expressions. All expressions are understood to be in normal coordinates unless otherwise noted.

Let us begin with deriving \cref{eqn:ChristoffelSymbol1}. As mentioned before, the $k=0$ case is understood to be just the same as \cref{eqn:SymGenChristoffelIsZero}. The $k=1$ case is fairly straightforward to see. Using the definition in \cref{eqn:GeneralChrisDef}, we find
\begin{equation}
\Gamma^\alpha_{\beta_1\cdots\beta_n\beta_{n+1}} = \Gamma^\alpha_{\beta_1\cdots\beta_n,\beta_{n+1}} - \sum_{j=1}^{n} \Gamma^\rho_{\beta_{n+1}\beta_j} \Gamma^\alpha_{\beta_1\cdots\rho\hat\beta_j\cdots\beta_n} \,.
\end{equation}
Evaluating it at the origin and symmetrize all the $\beta_i$ indices, we obtain
\begin{equation}
\overline\Gamma^\alpha_{(\beta_1\cdots\beta_n\beta_{n+1})} = \overline\Gamma^\alpha_{(\beta_1\cdots\beta_n,\beta_{n+1})} - \sum_{j=1}^{n} \overline\Gamma^\rho_{(\beta_{n+1}\beta_j} \overline\Gamma^\alpha_{\beta_1\cdots|\rho\hat\beta_j|\cdots\beta_n)} \,.
\end{equation}
In this equation, the term on the LHS vanishes due to \cref{eqn:SymGenChristoffelIsZero}, namely the $k=0$ case of \cref{eqn:ChristoffelSymbol1}. On the RHS, the second term also vanishes because its first factor, $\overline\Gamma^\rho_{\beta_{n+1}\beta_j}$, is also covered by the $k=0$ case of \cref{eqn:ChristoffelSymbol1}. Therefore, we conclude that the first term on the RHS must also vanish:
\begin{equation}
\overline\Gamma^\alpha_{(\beta_1\cdots\beta_n,\beta_{n+1})} = 0 \,.
\end{equation}

The above procedure ``$k=0 \;\Rightarrow\; k=1$'' generalizes to higher $k$ cases as well and \cref{eqn:ChristoffelSymbol1} can be proved by such an inductive procedure. To do so, let us now assume that \cref{eqn:ChristoffelSymbol1} holds for all $0\le k\le r$ already, and then we will show that it must also hold for $k=r+1$. We start with the $r$ partial derivatives on the LHS:
\begin{align}
\Gamma^\alpha_{\beta_1\cdots\beta_{n+1},\beta_{n+2}\cdots\beta_{n+r+1}}
&= \partial_{\beta_{n+2}} \cdots \partial_{\beta_{n+r+1}} \Gamma^\alpha_{\beta_1\cdots\beta_n\beta_{n+1}} \notag\\[5pt]
&\hspace{-60pt}
= \partial_{\beta_{n+2}} \cdots \partial_{\beta_{n+r+1}} \left( \Gamma^\alpha_{\beta_1\cdots\beta_n,\beta_{n+1}} - \sum_{j=1}^{n} \Gamma^\rho_{\beta_{n+1}\beta_j} \Gamma^\alpha_{\beta_1\cdots\rho\hat\beta_j\cdots\beta_n} \right) \notag\\[3pt]
&\hspace{-60pt}
= \Gamma^\alpha_{\beta_1\cdots\beta_n,\beta_{n+1}\cdots\beta_{n+r+1}} - \sum_{j=1}^{n} \partial_{\beta_{n+2}} \cdots \partial_{\beta_{n+r+1}} \left( \Gamma^\rho_{\beta_{n+1}\beta_j} \Gamma^\alpha_{\beta_1\cdots\rho\hat\beta_j\cdots\beta_n} \right) \,.
\label{eqn:GammaInduction}
\end{align}
The first term on the RHS has $r+1$ partial derivatives, which is the term of interest. In the second term, the $r$ partial derivatives need to be allocated onto the two Christoffel symbols, and in any of the resulting terms the first factor $\Gamma^\rho_{\beta_{n+1}\beta_j}$ can potentially get $0\le m \le r$ derivatives. The point is that upon symmetrizing all the $\beta_i$ indices and evaluating at the origin, they all vanish by our induction assumption, and so will the term on the LHS. Therefore, the first term on the RHS must also vanish:
\begin{equation}
\overline\Gamma^\alpha_{(\beta_1\cdots\beta_n,\beta_{n+1}\cdots\beta_{n+r+1})} = 0 \,.
\end{equation}
This completes our induction step and therefore proves \cref{eqn:ChristoffelSymbol1}.

Now with \cref{eqn:ChristoffelSymbol1}, we can easily derive \cref{eqn:ChristoffelSymbol2}. The relation
\begin{equation}
\Gamma_{\alpha\beta_1\beta_2} = g_{\alpha\lambda}\, \Gamma^\lambda_{\beta_1\beta_2} \,,
\end{equation}
gives us
\begin{equation}
\Gamma_{\alpha\beta_1\beta_2,\beta_3\cdots\beta_{k+2}} = \partial_{\beta_3}\cdots\partial_{\beta_{k+2}} \left( g_{\alpha\lambda}\, \Gamma^\lambda_{\beta_1\beta_2} \right) \,.
\end{equation}
Again, we need to allocate the $k$ partial derivatives onto the two factors in the parentheses. But in any of the resulting terms, we will have a factor of the Christoffel symbol $\Gamma^\lambda_{\beta_1\beta_2}$ with some number of partial derivatives appended. This factor vanishes once we symmetrize all the $\beta_i$ indices and evaluate it at the origin, because of (the $n=2$ case of) \cref{eqn:ChristoffelSymbol1}. Therefore, we obtain \cref{eqn:ChristoffelSymbol2}.

Now we can derive \cref{eqn:gsymmetrized}. The relation
\begin{equation}
2\, \Gamma_{\alpha\beta_1\beta_2} = g_{\alpha\beta_1,\beta_2} + g_{\alpha\beta_2,\beta_1} - g_{\beta_1\beta_2,\alpha} \,,
\end{equation}
obviously leads us to
\begin{equation}
2\, \Gamma_{\alpha\beta_1\beta_2,\beta_3\cdots\beta_{k+2}} = g_{\alpha\beta_1,\beta_2\beta_3\cdots\beta_{k+2}} + g_{\alpha\beta_2,\beta_1\beta_3\cdots\beta_{k+2}} - g_{\beta_1\beta_2,\beta_3\cdots\beta_{k+2}\alpha} \,.
\end{equation}
Symmetrizing all the $\beta_i$ indices and evaluating it at the origin, we find
\begin{equation}
2\, \overline\Gamma_{\alpha(\beta_1\beta_2,\beta_3\cdots\beta_{k+2})} = 2\,\overline{g}_{\alpha(\beta_1,\beta_2\beta_3\cdots\beta_{k+2})} - \overline{g}_{(\beta_1\beta_2,\beta_3\cdots\beta_{k+2})\alpha} \,.
\end{equation}
The LHS vanishes by \cref{eqn:ChristoffelSymbol2}, which implies that
\begin{equation}
2\,\overline{g}_{\alpha(\beta_1,\beta_2\beta_3\cdots\beta_{k+2})} - \overline{g}_{(\beta_1\beta_2,\beta_3\cdots\beta_{k+2})\alpha} = 0 \,.
\label{eqn:gk3sum}
\end{equation}
Note that the $k=0$ case of this is nothing but \cref{eqn:g3sum}. Following the same logic from \cref{eqn:g3sum} to \cref{eqn:gFirstDerivativeZero}, we first symmetrize all the indices in \cref{eqn:gk3sum} to obtain
\begin{equation}
\overline{g}_{(\alpha\beta_1,\beta_2\beta_3\cdots\beta_{k+2})} = 0 \,.
\end{equation}
On the other hand, upon expansion this quantity yields
\begin{equation}
0 = \overline{g}_{(\alpha\beta_1,\beta_2\beta_3\cdots\beta_{k+2})}
= \frac{1}{k+3} \left[ 2\, \overline{g}_{\alpha(\beta_1,\beta_2\beta_3\cdots\beta_{k+2})}
+ (k+1)\, \overline{g}_{(\beta_1\beta_2,\beta_3\cdots\beta_{k+2})\alpha} \right] \,.
\end{equation}
Combining this with \cref{eqn:gk3sum}, we get \cref{eqn:gsymmetrized}.

Finally, let us derive \cref{eqn:VNormalCoords}. For any $n+1$ ($n\ge0$) covariant derivatives of the potential, \ie, $V_{;\beta_1\cdots\beta_{n+1}}$, let us consider its $r^\text{th}$ ($r\ge0$) partial derivatives:
\begin{align}
V_{;\beta_1\cdots\beta_{n+1},\beta_{n+2}\cdots\beta_{n+r+1}} &= \partial_{\beta_{n+2}} \cdots \partial_{\beta_{n+r+1}} V_{;\beta_1\cdots\beta_{n+1}} \notag\\[8pt]
&\hspace{-60pt}
= \partial_{\beta_{n+2}} \cdots \partial_{\beta_{n+r+1}} \left( V_{;\beta_1\cdots\beta_n,\beta_{n+1}} - \sum_{j=1}^n \Gamma^\rho_{\beta_{n+1}\beta_j} V_{;\beta_1\cdots\rho\hat\beta_j\cdots\beta_n} \right) \notag\\[5pt]
&\hspace{-60pt}
= V_{;\beta_1\cdots\beta_n,\beta_{n+1}\cdots\beta_{n+r+1}} - \sum_{j=1}^n \partial_{\beta_{n+2}} \cdots \partial_{\beta_{n+r+1}} \left( \Gamma^\rho_{\beta_{n+1}\beta_j} V_{;\beta_1\cdots\rho\hat\beta_j\cdots\beta_n} \right) \,.
\end{align}
Similar to \cref{eqn:GammaInduction}, the second term above vanishes once we symmetrize all the $\beta_i$ indices and evaluate it at the origin, because of (the $n=2$ case of) \cref{eqn:ChristoffelSymbol1}. Therefore, we find
\begin{equation}
\overline{V}_{;(\beta_1\cdots\beta_{n+1},\beta_{n+2}\cdots\beta_{n+r+1})} = \overline{V}_{;(\beta_1\cdots\beta_n,\beta_{n+1}\cdots\beta_{n+r+1})} \,.
\end{equation}
This says that one can move the position of the comma towards the semicolon. Using this feature repeatedly, we can move the comma all the way:
\begin{equation}
\overline{V}_{;(\beta_1\cdots\beta_{n+1},\beta_{n+2}\cdots\beta_{n+r+1})} = \overline{V}_{;(\beta_1,\beta_2\cdots\beta_{n+r+1})} = \overline{V}_{,\beta_1\cdots\beta_{n+r+1}} \,,
\label{eqn:Vappended}
\end{equation}
where the second equation comes from the fact
\begin{equation}
V_{;\beta_1} = V_{,\beta_1}
\quad\Rightarrow\quad
V_{;\beta_1,\beta_2\cdots\beta_{n+r+1}} = V_{,\beta_1\cdots\beta_{n+r+1}} \,.
\end{equation}
Obviously, the $r=0$ case of \cref{eqn:Vappended} is just \cref{eqn:VNormalCoords}.

\pagebreak
\section{Perturbative Unitarity Bounds}
\label{sec:PerturbativeUnitarityBounds}

In this appendix, we review a formalism for deriving perturbative unitarity cutoffs from generic $n$-point amplitudes expounded in \cite{Abu-Ajamieh:2020yqi}.

We write the $S$-matrix as
\begin{equation}
S = I - i\, T \,,
\label{eqn:Tdef}
\end{equation}
and bracket it with incoming and outgoing multiparticle scattering states $|P,\alpha \rangle$ and $\langle Q, \beta |$, where $P,Q$ label the total momentum and $\alpha,\beta$ label all the other quantum numbers, such as angular momentum and particle species. To factor out overall momentum conservation, the states are normalized such that
\begin{equation}
\langle Q,\beta | P, \alpha \rangle = (2\pi)^4 \delta^{(4)}(Q-P) \delta_{\beta\alpha} \,.
\label{eq:StateNorm}
\end{equation}
Then, defining
\begin{subequations}
\begin{align}
\langle Q,\beta | S | P, \alpha \rangle &= (2\pi)^4 \delta^{(4)}(Q-P) \hat S_{\beta\alpha} \,, \\[5pt]
\langle Q,\beta | T | P, \alpha \rangle &= (2\pi)^4 \delta^{(4)}(Q-P) \hat M_{\beta\alpha} \,,
\end{align}
\end{subequations}
the $S$-matrix is rewritten
\begin{equation}
\hat S_{\beta\alpha} = \delta_{\beta\alpha} - i \hat M_{\beta\alpha} \, .
\end{equation}
The unitarity of $\hat S_{\beta \alpha}$ follows from the unitarity of the $S$-matrix, and leads to the optical theorem\footnote{Note that $\Im \hat M_{\alpha\alpha} \le 0$ with our convention of $T$ in \cref{eqn:Tdef}.}
\begin{equation}
0 = \sum_\beta | \hat S_{\beta \alpha}|^2 - 1 = 2\,\Im \hat M_{\alpha\alpha} + \sum_\beta | \hat M_{\beta \alpha}|^2 \,,
\label{eq:OpticalTheorem}
\end{equation}
where the $\sum_\beta$ is a sum over all states. Separating out the elastic process bounds the inelastic scattering
\begin{equation}
\sum_{\beta\neq\alpha} | \hat M_{\beta \alpha}|^2 = \sum_{\beta\neq\alpha} | \hat S_{\beta \alpha}|^2
= 1 - | \hat S_{\alpha \alpha}|^2 \le 1 \,.
\label{eq:PerturbativeUnitarity}
\end{equation}
This also implies an upper bound for any individual term
\begin{equation}
| \hat M_{\beta \alpha} |^2 \le 1 \qquad\forall\; \beta\ne\alpha \,.
\label{eq:PerturbativeUnitarityIndividual}
\end{equation}

In the following, we choose to project incoming and outgoing states onto their $s$-wave components, by averaging over incoming and outgoing phase space. Let $\alpha$ describe the $s$-wave component of an incoming $x$-particle state, which partitions $x=x_1 + \ldots + x_r$ into $r$ distinguishable particle species. Similarly, let $\beta$ be an outgoing $s$-wave state of $y=y_1 + \ldots + y_s$ particles, comprising $s$ distinguishable species. Then, in terms of the canonically normalized amplitude $\amp$,\footnote{Canonically normalized means constructed using single particle states $|p\rangle$ satisfying $\langle q | p \rangle = (2\pi)^3 \delta^{(3)}(\vec p - \vec q) \, 2 p^0$.}
\begin{align}
\hat M_{\beta\alpha} = \left( \frac{1}{ \left( \prod_{i=1}^r x_i! \right) \left( \prod_{j=1}^s y_j! \right) \mathrm{Vol}_x \mathrm{Vol}_y} \right)^\frac12 \int \mathrm{dLIPS}_x
\int \mathrm{dLIPS}_y\, \amp \,.
\label{eq:swaveproj}
\end{align}
The prefactor is fixed by the normalization condition in \cref{eq:StateNorm}, and is written in terms of the $n$-body phase space volume
\begin{equation}
\mathrm{Vol}_n \equiv \int \mathrm{dLIPS}_n = \int \left( \prod_{i=1}^n\frac{\dd^3 p_i}{(2\pi)^3\, 2 p^0_i} \right) (2\pi)^4 \delta^{(4)}\left( P - \sum_{i=1}^n p_i \right) \,.
\end{equation}
In the case of all massless particles, the phase space volume evaluates to \cite{Kleiss:1985gy},
\begin{equation}
\mathrm{Vol}_n = \frac{1}{8 \pi (n-1)! (n-2)!} \left( \frac{E}{4 \pi}  \right)^{2(n-2)} \,,
\label{eq:NBodyMasslessPhaseSpace}
\end{equation}
where $E=\sqrt{P^2}$ is the center-of-mass energy.

\section{Convergence Rate of Sectional Curvatures}
\label{appsec:an}

\begin{figure}[h!]
\centering
\includegraphics[width=0.45\textwidth]{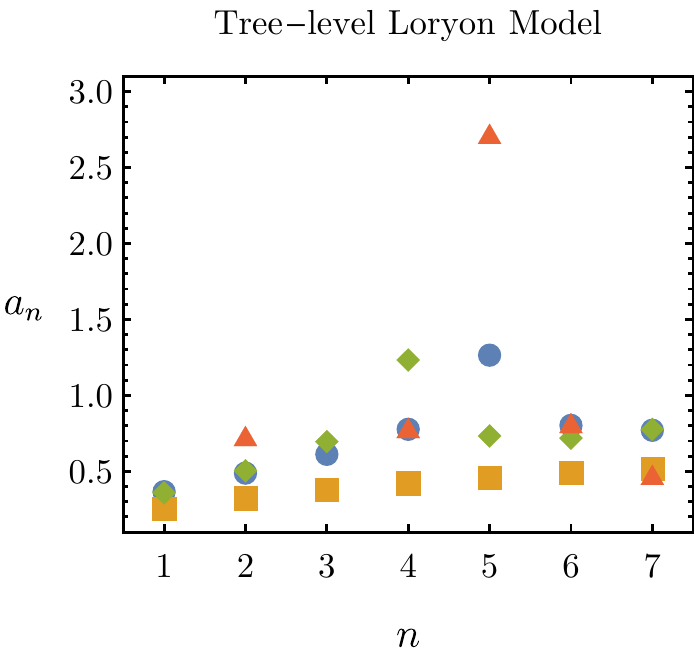}\hspace{40pt}\includegraphics[width=0.45\textwidth]{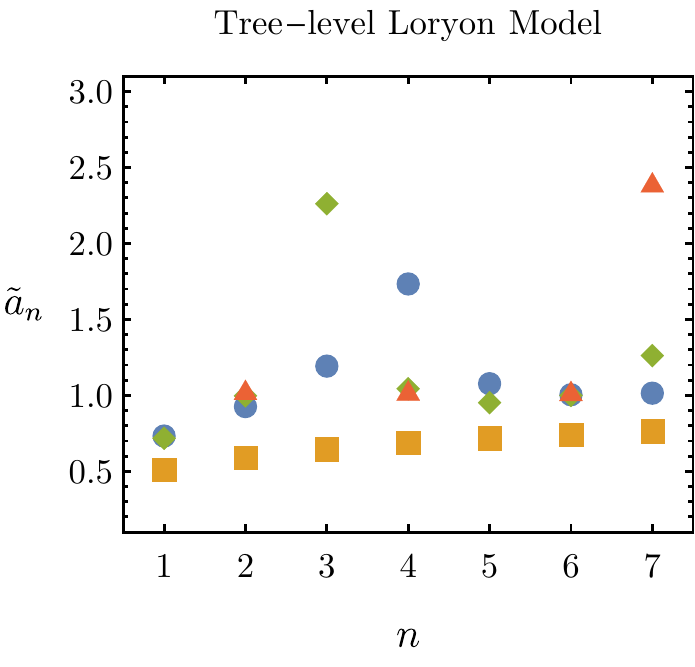}\\[2pt]
\includegraphics[width=0.45\textwidth]{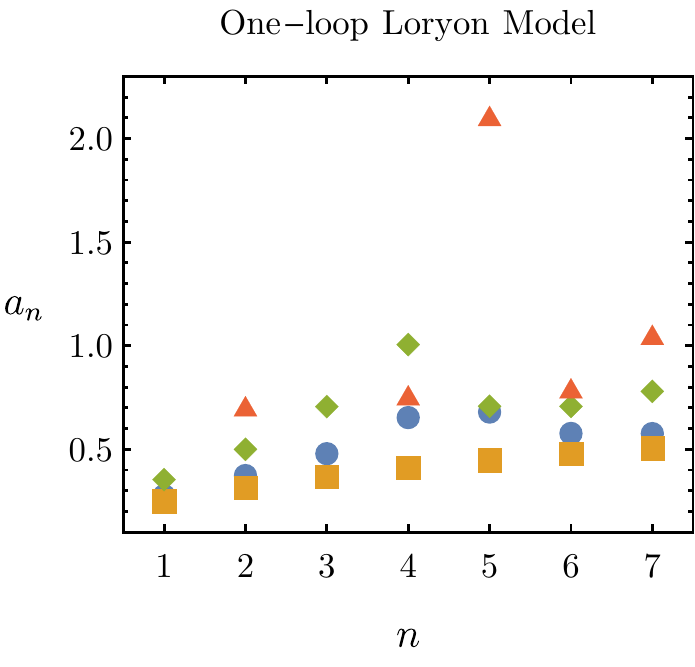}\hspace{40pt}\includegraphics[width=0.45\textwidth]{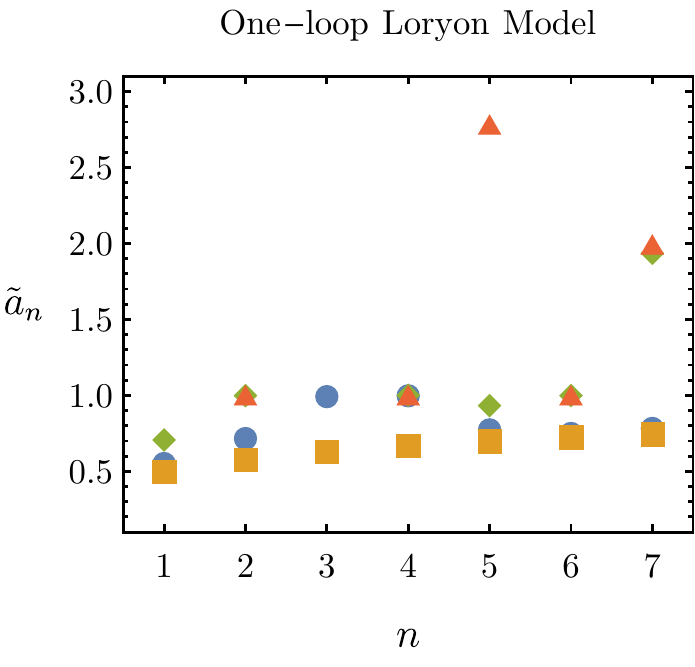}\\[2pt]
\hspace{-35pt}\includegraphics[width=0.45\textwidth]{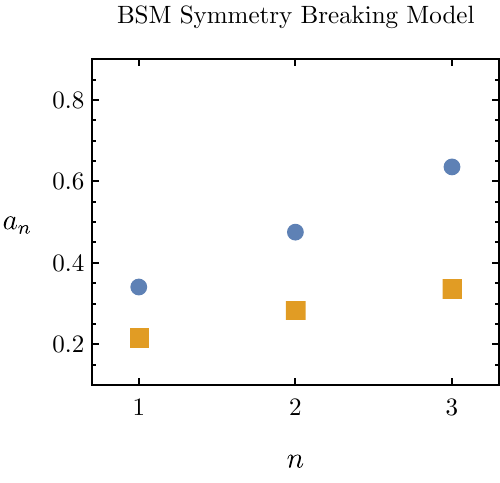}\\
   \includegraphics[align=c,width=0.26\textwidth]{figs/SingletLegend.pdf}
   \caption{These plots provide $a_n$ and $\tilde{a}_n$ (\cref{eq:appandef,eq:appantildedef}) as a function of $n$ for the benchmark points studied in \cref{sec:UVCompletions}. Note that $a_1(\tilde{a}_1)$ and $a_3(\tilde{a}_3)$ are outside the plot range for points D.}
  \label{fig:an}
\end{figure}

As discussed in \cref{sec:AmpGrowth}, the Cauchy-Hadamard theorem motivates us to relate the growth of derivatives of the sectional curvature to its radius of convergence by a quantity $a_n$ we defined in \cref{eq:LimitingScaleSectionalCurvature}. As $n\to \infty$, this quantity is expected to approach unity; see \cref{eq:CauchyHadamardInTermsOfan}. In this appendix, we provide numerical evidence that this is the case, by plotting
\begin{align}
  a_n &=  \left( \frac{v_\star^n}{n!} \frac{ |\partial_\canonh^n \mathcal{K}_h| }{|\mathcal{K}_h|} \right)^{\sfrac{-1}{n}} \bigg|_{\canonh=0} \,, \label{eq:appandef} \\[5pt]
  \tilde{a}_n &=  \left( \frac{v_\star^n}{n!} \frac{ |\partial_\canonh^n \mathcal{K}_\pi| }{|\mathcal{K}_\pi|} \right)^{\sfrac{-1}{n}} \bigg|_{\canonh=0} \,,\label{eq:appantildedef}
\end{align}
as a function of $n$ for the example model benchmarks introduced in \cref{sec:UVCompletions}, see \cref{fig:an}.

\end{spacing}

\clearpage
\newpage

\begin{spacing}{1.09}
\addcontentsline{toc}{section}{\protect\numberline{}References}%
\bibliographystyle{JHEP}
\bibliography{HEFTUnitarity}
\end{spacing}
\end{document}